\documentclass[12pt, draftclsnofoot, hidelinks, onecolumn]{IEEEtran}
\usepackage{epsfig,graphicx,subfigure,psfrag,amsmath,cases}
\usepackage{latexsym,amssymb,amsmath,epsfig,subfigure,algorithm,mathtools}
\usepackage{algorithmic}
\usepackage{color}
\usepackage{url}
\usepackage{scrtime}
\usepackage{cite}
\usepackage{epstopdf}
\usepackage{subfigure}
\usepackage{bbding}
\usepackage{multicol}
\usepackage{bm}
\usepackage{xcolor}
\usepackage[framemethod=TikZ]{mdframed}
\usetikzlibrary{shadows}
\usepackage{environ}
\usepackage{varwidth}

\newlength{\MyMdframedWidthTweak}%
\NewEnviron{MyMdframed}[1][]{%
    \setlength{\MyMdframedWidthTweak}{\dimexpr%
        +\mdflength{innerleftmargin}
        +\mdflength{innerrightmargin}
        +\mdflength{leftmargin}
        +\mdflength{rightmargin}
        }%
    \savebox0{%
        \begin{varwidth}{\dimexpr\linewidth-\MyMdframedWidthTweak\relax}%
            \BODY
        \end{varwidth}%
    }%
    \begin{mdframed}[
        backgroundcolor=lightgray,
        shadow=true,
        shadowsize=4pt,
        roundcorner=5pt,
        userdefinedwidth=\dimexpr\wd0+\MyMdframedWidthTweak\relax,
        #1]
        \usebox0
    \end{mdframed}
}

\author{Zhiqiang Wei, Weijie Yuan, Shuangyang Li, Jinhong Yuan, and Derrick Wing Kwan Ng\vspace{-18mm}
	\thanks{Zhiqiang Wei, Weijie Yuan, Shuangyang Li, Jinhong Yuan, and Derrick Wing Kwan Ng are with the School of Electrical Engineering and Telecommunications, the University of New South Wales, Australia (email: zhiqiang.wei; weijie.yuan; shuangyang.li; j.yuan; w.k.ng@unsw.edu.au).}}

\title{Transmitter and Receiver Window Designs for Orthogonal Time Frequency Space Modulation}

\newtheorem{T-Prob}{Transformed Problem}

\DeclareMathOperator{\Tr}{Tr}

\DeclareMathOperator{\mino}{minimize}
\DeclareMathOperator{\diag}{\mathrm{diag}}

\newcommand{\abs}[1]{\lvert#1\rvert}

\begin{document}
\maketitle
\begin{abstract}
In this paper, we investigate the impacts of transmitter and receiver windows on the performance of orthogonal time-frequency space (OTFS) modulation and propose window designs to improve the OTFS channel estimation and data detection performance.
In particular, assuming ideal pulse shaping filters at the transceiver, we derive the impacts of windowing on the effective channel and its estimation performance in the delay-Doppler (DD) domain, the total average transmit power, and the effective noise covariance matrix.
When the channel state information (CSI) is available at the transceiver, we analyze the minimum squared error (MSE) of data detection and propose an optimal transmitter window to minimize the detection MSE.
{The proposed optimal transmitter window is interpreted as a mercury/water filling power allocation scheme, where the mercury is firstly filled before pouring water to pre-equalize the TF domain channels.}
When the CSI is not available at the transmitter but can be estimated at the receiver, we propose to apply a Dolph-Chebyshev (DC) window at either the transmitter or the receiver, which can effectively enhance the sparsity of the effective channel in the DD domain.
Thanks to the enhanced DD domain channel sparsity, the channel spread due to the fractional Doppler is significantly reduced, which leads to a lower error floor in both channel estimation and data detection compared with that of rectangular window.
Simulation results verify the accuracy of the obtained analytical results and confirm the superiority of the proposed window designs in improving the channel estimation and data detection performance over the conventional rectangular window design.
\end{abstract}

\vspace{-4mm}
\section{Introduction}


Future wireless networks are expected to provide high-speed and ultra-reliable communications for a wide range of emerging mobile applications\cite{Hadani2017orthogonal}, including online video gaming, vehicle-to-vehicle (V2V),  vehicle-to-everything (V2X), high-speed railway systems, etc.
In practice, communications in high mobility scenarios suffer from severe Doppler spread, which deteriorates the performance of the widely adopted orthogonal frequency division multiplexing (OFDM) modulation in the current  fourth-generation (4G) and the emerging fifth-generation (5G) networks.
%
%
%
%
Recently, an increasing amount of attention has been paid for designing new modulation waveforms and schemes to meet the challenging requirements of high mobility communications for the next generation of wireless networks.

The wireless channel in high mobility propagation environments is inherently a linear time-variant fading channel\cite{BelloLTVC}, instead of the commonly assumed linear time-invariant one for each OFDM symbol as assumed in most of current communication systems \cite{DerrickEEOFDMA}.
In particular, the multipath propagation and the temporal channel variations give rise to the frequency-selective fading (time dispersion) and time-selective fading (frequency dispersion), respectively, resulting in the so called \textit{doubly-selective} or \textit{doubly-dispersive} channels\cite{Sayeed1999Joint,Ma2003Maximum}.
It is well known that OFDM is nearly capacity-achieving in time-invariant frequency-selective channels via applying an appropriate power and rate allocation as well as powerful modern error correcting codes\cite{Tse2005}.
%
In addition, OFDM is essentially a multi-carrier modulation scheme transferring a frequency-selective fading channel to multiple parallel frequency-flat subchannels, which facilitate the application of efficient channel estimation and equalization in the frequency domain.
Another benefit is its capability to combat the inter-symbol interference (ISI) with proper guard intervals inserted in the time domain since the OFDM symbol duration has been enlarged significantly compared to that of single-carrier systems.
In time-varying channels\cite{BelloLTVC}, the orthogonality promised in OFDM signals breaks down due to the power leakage among subcarriers, i.e., inter-carrier interference (ICI), induced by inevitable Doppler spread\cite{WangOFDMDoppler}.
As a result, the performance of both channel estimation and data detection of OFDM systems degrades dramatically in time-varying channels when conventional wireless transceivers are adopted.

\begin{figure}[t]
	\centering\vspace{-3mm}
	\includegraphics[width=5in]{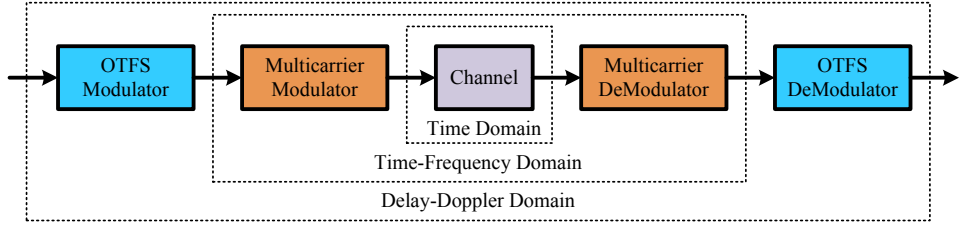}\vspace{-7mm}
	\caption{The concept of OTFS modulation and demodulation.}\vspace{-10mm}
	\label{OTFS}%
\end{figure}

Recently, a new two-dimensional (2D) modulation scheme referred to as the orthogonal time-frequency space (OTFS) modulation was proposed in \cite{Hadani2017orthogonal}, where the data symbols are multiplexed in the delay-Doppler (DD) domain rather than in the time-frequency (TF) domain as the traditional OFDM modulation\cite{BiglieriERROR,RavitejaEffectiveDiversity}.
As shown in Fig. \ref{OTFS}, an OTFS modulator transforms data symbols from the DD domain to the TF domain via a set of 2D orthogonal basis functions, which span across a time and frequency grid corresponding to the resource blocks available for an OTFS frame.
Then, a multi-carrier modulator, such as OFDM, is employed to further transform the OTFS modulated signals from the TF domain to the time domain for transmission over wireless channels.
At the receiver side, a cascade of a multi-carrier demodulator and an OTFS demodulator transforms the received signals back to the DD domain and then retrieves the transmitted data symbols.
In fact, OTFS modulation can be implemented based on conventional OFDM \cite{FarhangCPOFDM,RezazadehReyhaniCPOFDM} and pulse-shaped OFDM \cite{RavitejaOTFS} transceiver structures by simply adding pre-processing and post-processing blocks, which makes it attractive from a practical implementation point of view.

OTFS modulation is promising for high mobility communications.
By exploiting the basis expansion model (BEM)\cite{GiannakisBEM}, OTFS modulation effectively transforms the TF domain time-variant channel into an effective 2D time-invariant channel in the DD domain, which exhibits a sparse and stable property\cite{Hadani2017orthogonal,RavitejaOTFS}.
Besides, OTFS directly exploits the multipath fading and channel fluctuations via the DD domain multiplexing, which is inherently suitable to tackle the dynamics of time-variant channels, compared to the conventional OFDM technique.
In particular, it has been demonstrated that OTFS is resilient to delay-Doppler shifts and outperforms OFDM significantly for both uncoded \cite{RavitejaOTFS} and coded \cite{Hadani2017orthogonal,ZemenOrthogonalPrecoding,ShuangyangOTFS} systems.
%
%
More importantly, the 2D transformation from the DD domain to the TF domain employed by an OTFS modulator allows each information symbol to experience the whole TF domain channel over an OTFS frame.
Thus, OTFS enjoys the \textit{joint time-frequency diversity}\cite{SurabhiDiversityAnalysisBER} (the so-called \textit{full diversity} in \cite{Hadani2017orthogonal}), which is desirable to provide reliable communications over doubly dispersive channels.
Apart from the advantages of potentially exploiting full diversity and Doppler-resilience, some interesting side benefits of OTFS have also been revealed by researchers recently.
For instance, OTFS results in a lower peak-to-average power ratio (PAPR) compared to that of OFDM and generalized frequency division multiplexing (GFDM)\cite{SurabhiPAPROTFS}, which enables a high power transmission efficiency.
Additionally, as OTFS is essentially a block-wise transmission and detection scheme, it can reduce the cyclic prefix overhead with a proper pulse shaping filter design compared to OFDM systems\cite{RavitejaPulseShapingFilterOTFS}.
Furthermore, it has been demonstrated that OTFS is more robust against the carrier frequency offset between transceiver than OFDM\cite{SurabhiAsynchronousMMwAVE,ThajOTFSSDR}.
These advantages place OTFS in an ideal position for realizing high-mobility communication networks.

{However, as a new modulation scheme, OTFS introduces new critical challenges for transceiver architecture and algorithm designs, particularly for channel estimation and data detection in the presence of fractional inter-Doppler interference (IDI)\cite{RavitejaOTFS}.
Specifically, as derived in \cite{RavitejaOTFS}, the channel response and data symbols are coupled in the DD domain through a 2D circular convolution, where IDI is generally inevitable in the presence of multipath.
Moreover, IDI can be categorized as integer IDI and fractional IDI\cite{RavitejaOTFS}, depending on whether the Doppler shift indices are localized on the grid in the DD domain or not.
Most of existing works only considered integer IDI for simplicity, e.g. \cite{FarhangCPOFDM,RavitejaPulseShapingFilterOTFS,ShenOTFSMassiveMIMOCE}.
Yet, ensuring integer IDI requires a large speed separation among transceiver as well as all moving scatters to create a high Doppler resolution, which is not always possible in practical systems.
Therefore, this paper considers the commonly encountered case of fractional IDI for OTFS modulation, where the effective channel is spread across all the Doppler shift indices.
%
%
In practice, the channel estimation performance of OTFS systems is mainly limited by the IDI, where a guard space is usually required to avoid the IDI between data and pilot symbols, either employing a single pilot symbol \cite{RavitejaOTFSCE} or a pilot sequence\cite{ShenOTFSMassiveMIMOCE}.
Even worse, the IDI between data and pilot symbols caused by fractional Doppler becomes more severe leading to an error floor in the effective channel estimation\cite{OurTelstraReport}.
To lower the error floor, a much larger guard space between the data and pilot symbols is needed, which causes a higher amount of signaling overhead.
On the other hand, the more severe IDI among data symbols due to the fractional Doppler also imposes a challenge for realizing efficient data detection.
In fact, the computational complexity of the commonly adopted sum-product algorithm (SPA)-based data detection \cite{RavitejaOTFS} increases exponentially with the number of paths in effective channels in the DD domain.
Therefore, a pragmatic approach for reducing the channel spreading caused by fractional Doppler is needed.
Fortunately, as mentioned in \cite{FarhangCPOFDM,RezazadehReyhaniCPOFDM}, windowing in the TF domain has the potential in increasing the effective channel sparsity in the DD domain.
Yet, the authors in \cite{FarhangCPOFDM,RezazadehReyhaniCPOFDM} did not propose any method for window designs.
Besides, when the computational complexity of the data detection is not the bottleneck, the window designs also provide a new degrees of freedom to further improve the detection performance compared to the commonly used rectangular window\cite{RavitejaOTFS}.
In practice, the role of window in OTFS modulation and its impact on the performance of OTFS systems are not well understood.
To the best of our knowledge, there is no existing work studying on the window design for OTFS, which motivates this work.}

In this paper, we study the window design for OTFS modulation to improve the performance of channel estimation and data detection with the consideration of practical fractional IDI.
The main contributions of this work are given as follows:
\begin{itemize}
	\item We analyze the impact of windowing on OTFS systems, including the effective channel, the effective channel estimation performance, the average transmit power, and the effective noise covariance matrix.
	We find that the transmitter window (TX window) can be interpreted as a power allocation in the TF domain, while employing a receiver window (RX window) causes a colored noise.
	Also, we demonstrate that employing a window at either the transmitter or the receiver results in an identical error floor in the effective channel estimation.
	
	\item Considering the availability of the channel state information (CSI) at both the transmitter and the receiver, we analyze the data detection performance in terms of the mean squared error (MSE). 
	Then, we propose an optimal TX window design to minimize the detection MSE when the computational complexity is not the system limitation.
	It is interesting to point out that the proposed TX window design can be interpreted as a mercury/water filling power allocation scheme, where the mercury is firstly filled before pouring water to pre-equalize the eigen-channels in the TF domain.
	
	\item For a more practical case without CSI, i.e., CSI is not known at the transmitter but can be estimated at the receiver side, we propose to employ a Dolph-Chebyshev (DC) window in the TF domain.
	The employed DC window is optimal in the sense that it can obtain a predefined channel sparsity while suppressing the channel spreading caused by the fractional IDI to the largest degree.
	Due to the enhanced channel sparsity, applying the proposed DC window at either the transmitter or the receiver can achieve a much lower channel estimation error floor compared to the conventional rectangular window even with a smaller amount of guard space overhead.
	
	\item Extensive simulations are conducted to demonstrate the effectiveness of the proposed window designs for both cases of with and without CSI.
	In particular, with CSI, the proposed optimal TX window can achieve a significant detection performance improvement compared with the conventional rectangular window.
	For the case without CSI, the proposed DC window can achieve a substantial performance gain in terms of both channel estimation and data detection over the rectangular window.
	
\end{itemize}

%
%
%
%
%

{\textit{Notations:} Boldface capital and lower case letters are reserved for matrices and vectors, respectively; ${\left( \cdot \right)^{\mathrm{H}}}$ denotes the Hermitian transpose of a vector or matrix; $\mathbb{A}$ denotes the constellation set; $\mathbb{Z}^{+}$ denotes the set of all non-negative integers;
	$\mathbb{C}^{M\times N}$ denotes the set of all $M\times N$ matrices with complex entries;
	$\abs{\cdot}$ denotes the absolute value of a complex scalar;
	$\Tr\{\cdot\}$ denotes the trace operations; 	
	$E\{\cdot\}$ denotes the expectation;
	$\ast$ and $\circledast$ denote the convolution and circular convolution operations, respectively;
	$\left(\cdot\right)^*$ denotes the conjugate operation;
	$\otimes$ denotes the Kronecker product operator; $\left(\cdot\right)_N$ denotes the modulus operation with respect to $N$;
	$\diag\{\cdot\}$ returns a square diagonal matrix with the elements of input vector on the main diagonal;
	$\Re\{\cdot\}$ returns the real part of the input complex number;
	$\lfloor\cdot\rfloor$ is the floor function which returns the largest integer smaller than the input value;
	$\left[x\right]^+ = \max\left\{0,x\right\}$;
	$\mathbf{F}_N$ and $\mathbf{I}_M$ denote the discrete
	Fourier transform (DFT) matrix of size $N \times N$ and the identity matrix of size $M \times M$, respectively;
	The circularly symmetric complex Gaussian distribution with mean $\boldsymbol{\mu}$ and covariance matrix $\boldsymbol{\Sigma}$ is denoted by ${\cal CN}(\boldsymbol{\mu},\boldsymbol{\Sigma})$;
	$\sim$ stands for ``distributed as''.}

\begin{figure}[t]
	\centering\vspace{-5mm}
	\includegraphics[width=3.8in]{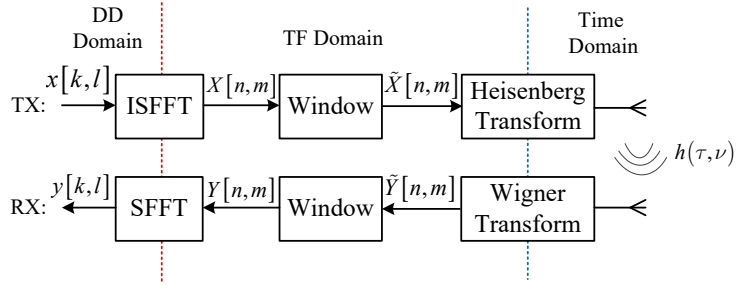}\vspace{-7mm}
	\caption{The block diagram of the OTFS transceiver.}\vspace{-10mm}
	\label{OFDM_OTFS}%
\end{figure}

\vspace{-4mm}
\section{System Model}
\vspace{-4mm}
\subsection{OTFS Transmitter}

A practical implementation of the OTFS transceiver is shown in Fig. \ref{OFDM_OTFS} \cite{RavitejaOTFS}.
{Without loss of generality, we assume that one OTFS frame occupies a bandwidth of $B_{\mathrm{OTFS}}$ and a time duration of $T_{\mathrm{OTFS}}$.
The total available bandwidth $B_{\mathrm{OTFS}}$ is divided into $M$ subcarriers with an equal spacing of $\Delta f = \frac{B_{\mathrm{OTFS}}}{M}$.
The total time duration $T_{\mathrm{OTFS}}$ is divided into $N$ time slots with an equal-length slot duration of $T = \frac{T_{\mathrm{OTFS}}}{N}$.
As a result, a grid of $N\times M$ can be constructed in the TF domain.
Note that the delay resolution is determined by the reciprocal of the system bandwidth, i.e., $\frac{1}{M \Delta f}$, while the Doppler resolution is determined by the OTFS frame duration, i.e., $\frac{1}{NT}$ \cite{RavitejaOTFS}.
Correspondingly, in the DD domain, $N$ denotes the number of Doppler indices with a Doppler resolution of $\frac{1}{NT}$ and $M$ denotes the number of delay indices with a delay resolution of $\frac{1}{M \Delta f}$.}
Consider a baseband modulated symbol in the DD domain:
\vspace{-2mm}
\begin{equation} \label{DD_DomainSymbol}
x\left[ {k,l} \right] \in \mathbb{A} = \{a_1,\ldots,a_Q\},\vspace{-2mm}
\end{equation}
where $k \in \{0,\ldots,N-1\}$ represents the Doppler index, $l \in \{0,\ldots,M-1\}$ represents the delay index, and $\mathbb{A}$ denotes the constellation set with a size of $Q$.
We assume that a normalized constellation is adopted, i.e., $E\left\{ {{{\left| {x\left[ {k,l} \right]} \right|}^2}} \right\} = 1$, and a proper scrambler is applied to scramble the output of the encoder such that it is reasonable to assume $E\left\{ {{{\left| {x\left[ {k,l} \right]} \right|}}} {{{\left| {x\left[ {k',l'} \right]} \right|}}}\right\} = 0$, $\forall k\neq k'$, $\forall l\neq l'$.
%
%
OTFS modulator performs a 2D transformation which maps the data symbols $x\left[ {k,l} \right]$ in the DD domain to $X\left[ {n,m} \right]$ in the TF domain.
In particular, such mapping can be realized by the inverse symplectic finite Fourier transform (ISFFT)\cite{Hadani2017orthogonal}:
\vspace{-2mm}
\begin{equation} \label{OTFS_Mod}
X\left[ {n,m} \right] = \frac{1}{{\sqrt {NM} }}\sum\limits_{k = 0}^{N - 1} {\sum\limits_{l = 0}^{M - 1} {x\left[ {k,l} \right]{e^{j2\pi \left( {\frac{{nk}}{N} - \frac{{ml}}{M}} \right)}}} },\vspace{-2mm}
\end{equation}
where $n \in \{0,\ldots,N-1\}$ is the time index and $m \in \{0,\ldots,M-1\}$ is the subcarrier index.
%
%
%
We note that the OTFS modulator can also been viewed as a 2D spreading/precoding from the DD domain to the TF domain, where each DD domain symbol $x\left[ {k,l} \right]$ is spread by a matrix (2D sequence) in the TF domain.

A TF domain TX window $U\left[ {n,m} \right]$ can be imposed through a point-wise multiplication with the TF domain signal $X\left[ {n,m} \right]$:
\vspace{-3mm}
\begin{equation} \label{TF_WindowTx}
\widetilde{X}\left[ {n,m} \right] = U\left[ {n,m} \right]X\left[ {n,m} \right],\vspace{-2mm}
\end{equation}
where $U\left[ {n,m} \right] \in \mathbb{C}$ denotes the TX window on the point of $\left[ {n,m} \right]$ in the TF domain grid.
Since a point-wise multiplication in the TF domain is equivalent to a 2D circular convolution in the DD domain\cite{RavitejaOTFS}, the TF domain window can be designed as a DD domain filter which can be exploited to improve the sparsity of the effective channel in the DD domain\cite{Hadani2017orthogonal}.
Then, a multicarrier modulator is utilized to transform the TF domain signal $\widetilde{X}\left[ {n,m} \right]$ to a time-domain signal $s\left( t \right)$, given by
\vspace{-2mm}
\begin{equation} \label{MulticarrierModI}
s\left( t \right) = \sum\limits_{n = 0}^{N - 1} {\sum\limits_{m = 0}^{M - 1} {\widetilde X\left[ {n,m} \right]{{g_{{\rm{tx}}}}\left( {t - nT} \right){e^{j2\pi m\Delta f\left( {t - nT} \right)}}} }},\vspace{-2mm}
\end{equation}
which is referred to as the Heisenberg transform in \cite{Hadani2017orthogonal}, where $t$ denotes the continuous time variable.
The time domain function ${{g_{{\rm{tx}}}}\left( {t} \right)}$ is the pulse-shaping filter of the multicarrier modulator for the windowed TF domain symbol.
To maintain the orthogonality of multicarrier modulation, we need to satisfy that $T\Delta f =1$ \cite{RavitejaOTFS} and the pulse shaping filter has a time duration of $T$, i.e., ${g_{{\rm{tx}}}}\left( t \right) = 0, t < 0, t> T$.
In this case, we have
\vspace{-2mm}
\begin{equation} \label{OrthogonalityMulticarrierMod}
\int_0^T {{g_{{\rm{tx}}}}\left( {t - nT} \right){e^{j2\pi m\Delta f\left( {t - nT} \right)}}g_{{\rm{tx}}}^ * \left( {t - n'T} \right){e^{ - j2\pi m'\Delta f\left( {t - n'T} \right)}}dt} = \delta \left[ {n-n'} \right]\delta \left[ {m-m'} \right],\vspace{-2mm}
\end{equation}
where $\delta \left[ \cdot \right]$ is the Dirac delta function, $n' \in \{0,\ldots,N-1\}$ and $m' \in \{0,\ldots,M-1\}$ represent the time and subcarrier indices, respectively.
In the next section, we will discuss the channel model in the DD domain.

\vspace{-4mm}
\subsection{DD Domain Channel Response}
For a linear time-variant channel, the received signal in the time domain is given by\cite{RavitejaOTFS}
\vspace{-2mm}
\begin{equation}\label{LTVChannelII}
r\left( t \right) = \int {\int {h\left( {\tau ,\nu } \right)} } {{e^{j2\pi \nu \left( {t - \tau } \right)}}}s\left( {t - \tau } \right)d\tau d\nu + w\left(t\right),\vspace{-2mm}
\end{equation}
where $w\left(t\right)$ denotes the noise signal in the time domain following a stationary Gaussian random process and we have $w\left(t\right) \sim \mathcal{CN}\left(0,N_0\right)$.
In practice, only few reflectors are moving within one OTFS frame duration and thus only a small number of channel taps are associated with Doppler shift\cite{Hadani2017orthogonal,RavitejaOTFS}.
Therefore, the resulting channel response in the DD domain is sparse compared to the whole DD domain grid spanned by one OTFS frame.
In particular, considering a channel consisting of $P$ independent distinguishable paths, the channel response in the DD domain can be modeled by
\vspace{-2mm}
\begin{equation}\label{LTVChannelDD}
{h\left( {\tau ,\nu } \right)} = \sum_{i=1}^{P} {h_i} \delta(\tau-\tau_i)\delta(\nu-\nu_i),\vspace{-2mm}
\end{equation}
where $h_i$, $\tau_i$, and $\nu_i$ denote the channel coefficient, delay, and Doppler shift associated with the $i$-th path, respectively.

When the delay and Doppler resolutions are sufficient for separating $\tau_i$ and $\nu_i$ in different paths in \eqref{LTVChannelDD}, i.e.,
\vspace{-2mm}
\begin{equation} \label{DDIntegerIDI}
\tau_i = l_{\tau_i}\frac{1}{M\Delta f} \;\;\;\text{and} \;\;\;
\nu_i = k_{\nu_i}\frac{1}{N T}, l_{\tau_i}, k_{\nu_i} \in \mathbb{Z},\vspace{-2mm}
\end{equation}
the channel response is perfectly localized on the grid of the DD domain.
In practice, the available system bandwidth in current cellular systems is usually sufficient to separate multiple paths' delay\cite{RavitejaOTFS}.
However, it might not be sufficient to resolve different Doppler shifts as a longer OTFS frame duration incurs a longer latency.
In particular, for an OTFS communication system with a carrier frequency $f_c$, bandwidth $B_{\mathrm{OTFS}} = M \Delta f$, and frame length $T_{\mathrm{OTFS}} = NT$, the Doppler resolution is given by $\frac{1}{NT} = \frac{B_{\mathrm{OTFS}}}{NM}$ which corresponds to a latency of $T_{\mathrm{OTFS}}$ and a speed resolution of $\Delta v = \frac{1}{T_{\mathrm{OTFS}}} \frac{c}{f_c}$, where $c$ is the speed of light.
For instance, for a microwave communication system with $f_c = 3$ GHz, $B_{\mathrm{OTFS}} = 15$ MHz, $M = 1024$, and $N = 16$, the latency $T_{\mathrm{OTFS}} = \frac{NM}{B_{\mathrm{OTFS}}} \approx 1.1$ ms is acceptable, while the corresponding speed resolution\footnote{Here, the speed resolution denotes the difference between the relative speeds among transceiver and  scatters causing Doppler shifts.} for ensuring integer Doppler is $\Delta v \approx 91.55\;\text{m/s} = 329.58\;\text{km/h}$.
However, this speed difference may not always be satisfied in practice.
%
%
Therefore, in contrast to the existing works focusing on only integer Doppler, it is necessary to consider the case with fractional Doppler, i.e.,
\vspace{-2mm}
\begin{equation} \label{DDFractionalIDI}
\nu_i = \left(k_{\nu_i} + \kappa_{\nu_i}\right)\frac{1}{N T},\vspace{-2mm}
\end{equation}
where $-\frac{1}{2} < \kappa_{\nu_i} < \frac{1}{2}$.

\vspace{-6mm}
\subsection{OTFS Receiver}
At the receiver side, we first perform a multicarrier demodulation for the received signal $r\left( t \right)$ with a receiving filter to obtain the TF domain signal $\widetilde{Y}\left[n,m\right]$, given by:
\vspace{-2mm}
\begin{equation}\label{MulticarrierDeMod}
\widetilde{Y}\left[n,m\right] = \int {r\left( t \right)g_{{\rm{rx}}}^ * \left( {t - nT} \right){e^{ - j2\pi m\Delta f\left( {t - nT} \right)}}dt},\vspace{-2mm}
\end{equation}
which is referred to as the Wigner transform in \cite{Hadani2017orthogonal}.
A time domain function ${{g_{{\rm{rx}}}}\left( {t} \right)}$ serves as a receiving filter for the multicarrier demodulator to sample the discrete symbol $\widetilde Y\left[ {n,m} \right]$ from the received waveform $r\left( t \right)$, which should be designed corresponding to the transmit pulse shaping filter in \eqref{MulticarrierModI}.
Substituting \eqref{MulticarrierModI}, \eqref{LTVChannelII}, and \eqref{LTVChannelDD} into \eqref{MulticarrierDeMod}, we have\cite{RavitejaOTFS}
\vspace{-2mm}
\begin{equation} \label{TFIOFelationship}
\widetilde{Y}\left(n,m\right) = \sum\limits_{n' = 0}^{N - 1} {\sum\limits_{m' = 0}^{M - 1} {\widetilde X\left[ {n',m'} \right]} } \widetilde H_{n,m}\left[n',m'\right] + \widetilde{Z}\left[n,m\right],\vspace{-2mm}
\end{equation}
where $\widetilde{Z}\left[n,m\right] \sim \mathcal{CN}\left(0,N_0\right)$ denotes the additive white Gaussian noise (AWGN) samples in the TF domain. Besides, the TF domain effective channel is given by
\vspace{-2mm}
\begin{equation}
\widetilde H_{n,m}\left[n',m'\right] = \sum_{i=1}^{P} {h_i} {{A_{{g_{\rm{rx}}}{g_{\rm{tx}}}}}\left( {\left( {n - n'} \right)T - \tau_i ,\left( {m - m'} \right)\Delta f - \nu_i } \right)}
{e^{j2\pi \nu_i nT}}{e^{ - j2\pi \left( {m'\Delta f + \nu_i } \right)\tau_i }}\vspace{-2mm}
\end{equation}
where ${{A_{{g_{\rm{rx}}}{g_{\rm{tx}}}}}\left(\tau,\nu \right)}$ is the cross-ambiguity function between  ${g_{{\rm{tx}}}}\left( {t  } \right)$ and ${g_{{\rm{rx}}}}\left( {t  } \right)$ \cite{RavitejaOTFS}, given by
\vspace{-2mm}
\begin{equation} \label{AmbiguityFunction}
{{A_{{g_{\rm{rx}}}{g_{\rm{tx}}}}}\left(\tau,\nu \right)} = \int_{t} {g_{{\rm{tx}}}}\left( {t} \right)g_{{\rm{rx}}}^ * \left( {t - \tau } \right){e^{ - j2\pi \nu \left( {t - \tau } \right)}}dt.\vspace{-2mm}
\end{equation}
We can observe that ISI exists when $\widetilde H_{n,m}\left[n',m'\right] \neq 0$, $\forall n \neq n'$, and ICI retains when $\widetilde H_{n,m}\left[n',m'\right] \neq 0$, $\forall m \neq m'$, in the TF domain.

The transceiver pulse shaping filters, ${g_{{\rm{tx}}}}\left( {t  } \right)$ and ${g_{{\rm{rx}}}}\left( {t  } \right)$, are said to be ideal if they satisfy the bi-orthogonal condition\cite{RavitejaOTFS}:
\vspace{-2mm}
\begin{equation}\label{BIOFthogonal}
{{A_{{g_{rx}}{g_{tx}}}}\left( {\left( {n - n'} \right)T - \tau_i ,\left( {m - m'} \right)\Delta f - \nu_i } \right)} = q_{\tau_{\mathrm{max}}}\left(t-nT\right)q_{\nu_{\mathrm{max}}}\left(f-m\Delta f\right),\vspace{-2mm}
\end{equation}
with
\vspace{-2mm}
\begin{equation}
{q_a}\left( {x - nb} \right) = \left\{ {\begin{array}{*{20}{c}}
	{\delta \left[ n \right]}&{\left| {x - nb} \right| \le a},\\[-1mm]
	{q\left( x \right)}&{\mathrm{otherwise}},
	\end{array}} \right.\vspace{-2mm}
\end{equation}
where $q\left( x \right)$ is an arbitrary function.
In this work, to facilitate the window design, we assume that the ideal transceiver pulse shaping filters\footnote{As proved in \cite{RavitejaOTFS}, the locations of the non-zero entries in the effective channel in the DD domain are identical for both ideal and rectangular pulse shaping filters with only an additional phase difference. As a result, the IDIs also share the same interference pattern in the DD domain. Therefore, the proposed window design adopting ideal pulse shaping filters can be straightforwardly extended to the case of the commonly adopted rectangular pulse.} satisfy the bi-orthogonal condition in \eqref{BIOFthogonal}, i.e., ISI and ICI free in the TF domain, and the received signal in the TF domain can be given by
\vspace{-2mm}
\begin{equation} \label{TFIOIdealPulse}
\widetilde{Y}\left(n,m\right) = {\widetilde X\left[ {n,m} \right]} \widetilde H\left[n,m\right]  + \widetilde{Z}\left[n,m\right],\vspace{-2mm}
\end{equation}
where the TF domain effective channel is given by
\vspace{-2mm}
\begin{equation} \label{TFIOChannelIdealPulse}
\widetilde H\left[n,m\right] =\sum\limits_{i = 1}^{P}  {h_i {e^{ - j2\pi \frac{\left(k_{\nu_i} + \kappa_{\nu_i}\right)l_{\tau_i}}{NM}}} {e^{j2\pi \left( {\frac{{n\left(k_{\nu_i} + \kappa_{\nu_i}\right)}}{N} - \frac{{ml_{\tau_i}}}{M}} \right)}}  } .\vspace{-2mm}
\end{equation}

Corresponding to the TX window, we can insert a RX window $V\left[ {n,m} \right]$ to the received signal in the TF domain:
\vspace{-2mm}
\begin{equation} \label{TF_WindowRx}
{Y}\left[ {n,m} \right] = V\left[ {n,m} \right]\widetilde{Y}\left[ {n,m} \right],\vspace{-2mm}
\end{equation}
where $V\left[ {n,m} \right] \in \mathbb{C}$.
Then, an OTFS demodulator transforms the TF domain signals ${Y}\left[ {n,m} \right]$ to the DD domain signals $y\left[ {k,l} \right]$ through a symplectic finite Fourier transform (SFFT) \cite{Hadani2017orthogonal}:
\vspace{-2mm}
\begin{equation} \label{OTFS_DeMod}
y\left[ {k,l} \right] = \frac{1}{{\sqrt {NM} }}\sum\limits_{n = 0}^{N - 1} {\sum\limits_{m = 0}^{M - 1} {Y\left[ {n,m} \right]{e^{-j2\pi \left( {\frac{{kn}}{N} - \frac{{lm}}{M}} \right)}}} }.\vspace{-2mm}
\end{equation}

\vspace{-4mm}
\subsection{Equivalent Vectorization Form}
Let us define $\mathbf{x}_{\mathrm{DD}} \in \mathbb{C}^{MN \times 1}$ as the vectorization form of $x\left[ {k,l} \right]$ in \eqref{OTFS_Mod}, where its $\left(kM+l\right)$-th entry is $x\left[ {k,l} \right]$, and $\mathbf{y}_{\mathrm{DD}} \in \mathbb{C}^{MN \times 1}$ as the vectorization form of $y\left[ {k,l} \right]$ in \eqref{OTFS_DeMod}, where its $\left(kM+l\right)$-th entry is $y\left[ {k,l} \right]$.
Based on the OTFS transceiver structure introduced above, an equivalent vectorization form of the input-output relationship in the DD domain has been derived in \cite{FarhangCPOFDM,RezazadehReyhaniCPOFDM}, which is given as follow:
\vspace{-2mm}
\begin{align}\label{VectorForm}
\mathbf{y}_{\mathrm{DD}}
=& \left(  {{\bf{F}}_N} \otimes {\bf{F}}_M^{\rm{H}} \right) {\bf{V}} \left( {{{\bf{I}}_N} \otimes {{\bf{F}}_M}} \right) {{{{{\bf{H}}}_t}}}\left( { {{\bf{I}}_N} \otimes {\bf{F}}_M^{\rm{H}} } \right){\bf{U}}\left( { {\bf{F}}_N^{\rm{H}} \otimes {{\bf{F}}_M}} \right){{\bf{x}}_{{\rm{DD}}}} \notag\\[-1mm]
+& \left(\mathbf{F}_{N} \otimes \mathbf{F}_{M}^{\mathrm{H}} \right)\mathbf{V} \left(\mathbf{I}_N \otimes \mathbf{F}_{M} \right)\mathbf{w},
\end{align}
\vspace{-10mm}\par\noindent
where ${{\bf{F}}_M} \in \mathbb{C}^{M \times M}$ and ${{\bf{F}}_N} \in \mathbb{C}^{N \times N}$ denote the DFT matrices.
$\mathbf{U} \in \mathbb{C}^{NM \times NM}$ and $\mathbf{V} \in \mathbb{C}^{NM \times NM}$ are diagonal matrices, whose $\left(nM+m\right)$-th diagonal entry is $U\left[n,m\right]$ and $V\left[n,m\right]$, respectively.
The vector $\mathbf{w}$ is a sampled version of $w\left(t\right)$ at time $t = \left(nM+m\right)\frac{1}{M\Delta f}$ and thus it follows $\mathbf{w} \sim {\cal CN}(\mathbf{0},N_0\mathbf{I}_{MN})$.
The matrix ${{{{{\bf{H}}}_t}}}\in \mathbb{C}^{MN \times MN}$ is the time domain channel response including the effect of transceiver pulse shaping filters.
Note that the system model in \eqref{VectorForm} is applicable to any transceiver pulse shaping filter.
The TF domain channel matrix for the vectorized channel input ${\widetilde{\bf{x}}_{{\rm{TF}}}} = {\bf{U}}\left( { {\bf{F}}_N^{\rm{H}} \otimes {{\bf{F}}_M}} \right){{\bf{x}}_{{\rm{DD}}}}$ and output ${\widetilde{\bf{y}}_{{\rm{TF}}}} = \left( {{{\bf{I}}_N} \otimes {{\bf{F}}_M}} \right) {{{{{\bf{H}}}_t}}}\left( { {{\bf{I}}_N} \otimes {\bf{F}}_M^{\rm{H}} } \right){\bf{U}}\left( { {\bf{F}}_N^{\rm{H}} \otimes {{\bf{F}}_M}} \right){{\bf{x}}_{{\rm{DD}}}}$ can be defined as
\vspace{-2mm}
\begin{equation}\label{TFDomainChannel}
\widetilde{\bf H}_{\rm TF} = \left(\textbf{I}_N \otimes \textbf{F}_M\right) {\bf H}_t \left(\textbf{I}_N \otimes \textbf{F}_M^{\rm H}\right) \in \mathbb{C}^{MN \times MN},\vspace{-2mm}
\end{equation}
which is a diagonal matrix when ideal pulse shaping filters are adopted at the transceivers.
Meanwhile, the DD domain channel matrix for the vectorized channel input ${{\bf{x}}_{{\rm{DD}}}}$ and output ${{\bf{y}}_{{\rm{DD}}}}$ can be obtained by
\vspace{-2mm}
\begin{equation}\label{DDDomainChannel}
	{{{\bf{H}}}_{{\rm{DD}}}} = {\left(  {{\bf{F}}_N} \otimes {\bf{F}}_M^{\rm{H}} \right) {{\bf{V}} {\left( {{{\bf{I}}_N} \otimes {{\bf{F}}_M}} \right) {{{{{\bf{H}}}_t}}}\left( { {{\bf{I}}_N} \otimes {\bf{F}}_M^{\rm{H}} } \right)}{\bf{U}}}\left( { {\bf{F}}_N^{\rm{H}} \otimes {{\bf{F}}_M}} \right)} \in \mathbb{C}^{MN \times MN}.\vspace{-2mm}
\end{equation}
The effective vectorized channel matrices in the TF domain and the DD domain will be utilized in the performance analysis of the detection MSE for OTFS.

\vspace{-4mm}
\section{The Impact of Windowing for OTFS Modulation}
In this section, we analyze the impacts of windowing for OTFS modulation on the effective channel, the effective channel estimation performance, the average transmit power, and the noise covariance matrix, which will serve as building blocks for practical window designs in the next section.

\vspace{-4mm}
\subsection{Effective Channel in the DD Domain}
Since the data symbols are multiplexed and detected in the DD domain, one might be interested to derive the input-output relationship via characterizing the effective channel in the DD domain.
According to the OTFS transceiver structure introduced above, the output of the OTFS demodulator in the DD domain is given by
\vspace{-2mm}
\begin{align} \label{OTFS_DeModIdealPulse}
y\left[ {k,l} \right] 
&=  \sum\limits_{k' = 0}^{N - 1} {\sum\limits_{l' = 0}^{M - 1} {x\left[ {k',l'} \right]} } {h_w}\left[ {k - k',l - l'} \right] + \sum\limits_{k' = 0}^{N - 1} {\sum\limits_{l' = 0}^{M - 1} {z\left[ {k',l'} \right]} } {v_z}\left[ {k - k',l - l'} \right],
\end{align}
\vspace{-8mm}\par\noindent
where ${h_w}\left[ {k ,l } \right]$ denotes the effective channel in the DD domain capturing the windows' effect and it is given by
\vspace{-2mm}
\begin{align} \label{DDIOChannelDiscreteIdealPulseFractional}
{h_w}\left[ {k ,l } \right]
&= \frac{1}{NM}\sum_{i=1}^{P} {h_i} \sum\limits_{n = 0}^{N - 1} \sum\limits_{m = 0}^{M - 1} {V\left[ {n,m} \right]U\left[ {n,m} \right]{e^{ - j2\pi n\frac{\left( {k-k_{\nu_i} - \kappa_{\nu_i}} \right)}{N}}}{e^{j2\pi m \frac{\left( l-l_{\tau_i} \right)}{M }}} {e^{ - j2\pi \frac{\left(k_{\nu_i} + \kappa_{\nu_i}\right)l_{\tau_i}}{NM} }}}\notag\\[-1mm]
&= \sum_{i=1}^{P} {h_i} w(k-k_{\nu_i}-\kappa_{\nu_i}, l-l_{\tau_i}) {e^{ - j2\pi \frac{\left(k_{\nu_i} + \kappa_{\nu_i}\right)l_{\tau_i}}{NM} }}.
\end{align}
\vspace{-8mm}\par\noindent
In \eqref{DDIOChannelDiscreteIdealPulseFractional}, $w(k - {k_{{\nu _i}}}-\kappa_{\nu_i},l - {l_{{\tau _i}}})$ is a DD domain filter designed by the joint TX-RX window and it is given by
\vspace{-2mm}
\begin{equation} \label{DDFilterIdealPulse}
w(k - {k_{{\nu _i}}}-\kappa_{\nu_i},l - {l_{{\tau _i}}}) = \frac{1}{NM}\sum\limits_{n = 0}^{N - 1} \sum\limits_{m = 0}^{M - 1} V\left[ {n,m} \right]U\left[ {n,m} \right]{e^{ - j2\pi n\frac{\left( {k-k_{\nu_i} - \kappa_{\nu_i}} \right)}{N}}}{e^{j2\pi m \frac{\left( l-l_{\tau_i} \right)}{M }}}.\vspace{-2mm}
\end{equation}
Also in \eqref{OTFS_DeModIdealPulse}, ${v_z}\left[ {k ,l } \right]$ is a DD domain filter induced by the RX window and is given by
\vspace{-2mm}
\begin{equation} \label{DDFilterRxIdealPulse}
{v_z}\left[ {k ,l } \right] = \frac{1}{NM}\sum\limits_{n = 0}^{N - 1} \sum\limits_{m = 0}^{M - 1} V\left[ {n,m} \right]{e^{ - j2\pi \frac{nk}{N}}}{e^{j2\pi \frac{ml}{M}}}.\vspace{-2mm}
\end{equation}

We can observe that different from the original DD domain channel response in \eqref{LTVChannelDD}, the effective channel in \eqref{DDIOChannelDiscreteIdealPulseFractional} has a circular structure due to ${h_w}\left[ {\left(k\right)_N ,\left(l\right)_M } \right] = {h_w}\left[ {k ,l } \right]$.
As such, from \eqref{OTFS_DeModIdealPulse}, we can observe that the received signal $y\left[ {k,l} \right]$ is a 2D circular convolution between the data symbols ${x\left[ {k,l} \right]}$ and the effective channel ${h_w}\left[ {k ,l } \right]$ in the DD domain.
Furthermore, as the data and training symbols are multiplexed in the DD domain \cite{RavitejaOTFSCE}, the channel estimation performance and the data detection complexity depend on the effective channel ${h_w}\left[ {k ,l } \right]$ instead of the original channel response $h\left( {\tau ,\nu } \right)$.
As shown in \eqref{DDIOChannelDiscreteIdealPulseFractional}, the effective channel ${h_w}\left[ {k ,l } \right]$ is a summation of the channel spread of each path where $w(k - {k_{{\nu _i}}}-\kappa_{\nu_i},l - {l_{{\tau _i}}}) \neq 0$, $\forall i$, and the spreading pattern can be manipulated by the design of the DD domain filter $w(k,l)$.
In other words, the channel sparsity of the effective channels can be controlled by the TX and RX windows.
In \eqref{DDFilterIdealPulse}, we can observe that imposing a TX window $U\left[ {n,m} \right]$ or a RX window $V\left[ {n,m} \right]$ in the TF domain has the same effect in designing the DD domain filter $w(k,l)$.
In contrast, only the RX window $V\left[ {n,m} \right]$ affects the DD domain filter, ${v_z}\left[ {k,l} \right] $, which alters the properties of the noise at the receiver side.

\vspace{-2mm}
\subsubsection{Effective Channel Sparsity with Rectangular Window}
Since the rectangular window is the most straightforward one to be considered\cite{RavitejaOTFS,RavitejaOTFSCE}, we investigate the  effective channel sparsity with rectangular window for both cases of integer and fractional Doppler.
When employing a rectangular window, i.e., $V\left[ {n,m} \right] = U\left[ {n,m} \right] =1$, $\forall n,m$, we have the DD domain filters given by
\vspace{-6mm}
\begin{align}
w(k - {k_{{\nu _i}}}-\kappa_{\nu_i},l - {l_{{\tau _i}}}) &= \mathcal{G}^{\mathrm{Rect}}_N\left({k - {k_{{\nu _i}}}  -\kappa_{\nu_i}}\right) \mathcal{F}^{\mathrm{Rect}}_M \left({l - {l_{{\tau _i}}}}\right)\; \text{and}\label{WindowDDDomainFractional}\\[-1mm]
{v_z}\left[ {k ,l } \right] &= \mathcal{G}^{\mathrm{Rect}}_N\left(k\right) \mathcal{F}^{\mathrm{Rect}}_M \left(l\right),\label{WindowDDDomainFractionalNoise}
\end{align}
\vspace{-8mm}\par\noindent
respectively.
Functions $\mathcal{G}^{\mathrm{Rect}}_N\left(k\right)$ and $\mathcal{F}^{\mathrm{Rect}}_M \left(l\right)$ represent the filters in the delay and Doppler domains, respectively, and they are given by
\vspace{-2mm}
\begin{equation}
\mathcal{G}^{\mathrm{Rect}}_N\left(k\right) = \frac{1}{N}\left({e^{ - j\left( {N - 1} \right) \frac{\pi k}{N}}}\frac{\sin \left(\pi k\right)}{{\sin \left( {\frac{{\pi k }}{N}} \right)}} \right) \;\text{and}\;
\mathcal{F}^{\mathrm{Rect}}_M \left(l\right) = \frac{1}{M}\left({{e^{ - j\left( {M - 1} \right) \frac{\pi l}{M}}}\frac{{\sin \left( {\pi l } \right)}}{{\sin \left( {\frac{{\pi l}}{M}} \right)}}} \right),\vspace{-2mm}
\end{equation}
respectively.

For the case of integer Doppler, i.e., $\kappa_{\nu_i} = 0$, the DD domain filter is simplified as
\vspace{-2mm}
\begin{equation}\label{WindowDDDomain}
w\left[k - {k_{{\nu _i}}},l - {l_{{\tau _i}}}\right] = \delta\left[k - {k_{{\nu _i}}}\right] \delta\left[l - {l_{{\tau _i}}}\right] = \left\{ {\begin{array}{*{20}{c}}
	{1}&{\left(k - {k_{{\nu _i}}}\right)_N = 0,\left(l - {l_{{\tau _i}}}\right)_M = 0}\\
	0& \mathrm{otherwise}
	\end{array}} \right.,\vspace{-2mm}
\end{equation}
and the effective channel in the DD domain is given by
\vspace{-2mm}
\begin{equation} \label{DDIOChannelDiscreteIdealPulseInteger}
{h_w}\left[ {k ,l } \right]
= \sum_{i=1}^{P} {h_i} \delta\left[k - {k_{{\nu _i}}}\right] \delta\left[l - {l_{{\tau _i}}}\right] {e^{ - j2\pi \frac{k_{\nu_i}l_{\tau_i}}{NM} }}.\vspace{-2mm}
\end{equation}
We can observe that the effective channel ${h_w}\left[ {k ,l } \right]$ in the DD domain has a response if and only if $k = {k_{{\nu _i}}}$ and $l = {l_{{\tau _i}}}$, i.e., the effective channel shares the same channel sparsity with the original DD domain channel response in \eqref{LTVChannelDD}.
Moreover, the effective channel ${h_w}\left[ {k ,l } \right]$ is a phase-rotated version of the original channel response in \eqref{LTVChannelDD}, where the delay and Doppler shift of the $i$-th path rotates the original channel response ${h_i}\in \mathbb{C}$ with a phase of $2\pi k_{\nu_i}l_{\tau_i} /MN$.

\begin{figure}[t]
	\centering\vspace{-5mm}
	\includegraphics[width=3.8in]{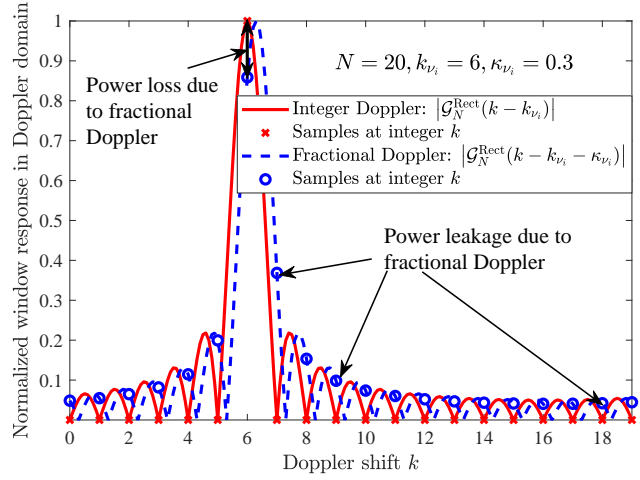}\vspace{-7mm}
	\caption{The channel spreading in the Doppler domain with a rectangular window with/without fractional Doppler, where $\mathrm{SL}_{w} \approx \frac{1}{N}$ denotes the sidelobe level of the adopted rectangular window.}\vspace{-10mm}
	\label{FractionalIDI}%
\end{figure}

For the case of fractional Doppler, i.e., $\kappa_{\nu_i} \neq 0$, the effective channel is given by
\vspace{-2mm}
\begin{equation} \label{DDIOChannelDiscreteIdealPulseFractional_RectWindow}
{h_w}\left[ {k ,l } \right]
= \sum_{i=1}^{P} {h_i} \mathcal{G}^{\mathrm{Rect}}_N\left({k - {k_{{\nu _i}}}  -\kappa_{\nu_i}}\right) \delta\left[l - {l_{{\tau _i}}}\right]\ {e^{ - j2\pi \frac{\left(k_{\nu_i} + \kappa_{\nu_i}\right)l_{\tau_i}}{NM} }}.\vspace{-2mm}
\end{equation}
From \eqref{DDIOChannelDiscreteIdealPulseFractional_RectWindow}, we can observe that the effective channel in the DD domain ${h_w}\left[ {k ,l } \right]$ has more ``paths'' (non-zero entries) than that of the original channel response in \eqref{LTVChannelDD} since the Doppler domain filter $\mathcal{G}^{\mathrm{Rect}}_N\left({k - {k_{{\nu _i}}}  -\kappa_{\nu_i}}\right) \neq 0$, $\forall k,{k_{{\nu _i}}}$, and $\forall \kappa_{\nu_i} \neq 0$.
In fact, for each path with a Doppler shift of ${k_{{\nu _i}}} + \kappa_{\nu_i}$, the channel coefficient ${h_i}$ is spread to all the Doppler indices $k$ in the Doppler domain.
To visualize the channel spreading, we ignore the delay domain at the moment and plot the Doppler domain filter response $\left|\mathcal{G}^{\mathrm{Rect}}_N\left({k - {k_{{\nu _i}}}  -\kappa_{\nu_i}}\right)\right|$ to illustrate the impact of  fractional Doppler in Fig. \ref{FractionalIDI}.
It can be seen that without the fractional Doppler, the filter $\left|\mathcal{G}^{\mathrm{Rect}}_N\left({k - {k_{{\nu _i}}}}\right)\right|$ is a perfect sampling function $\delta \left[k - {k_{{\nu _i}}}\right]$, i.e., no channel inter-spread.
However, the existence of the fractional Doppler shift $\kappa_{\nu_i}$ not only reduces signal power at the sampling point $k = {k_{{\nu _i}}}$ but also introduces non-negligible power leakage from the Doppler shift ${k_{{\nu _i}}}$ to $k \neq  {k_{{\nu _i}}}$.
In other words, with the application of the rectangular window, fractional Doppler sacrifices the sparsity of the effective channel in the DD domain, which could degrade the channel estimation performance and increase the complexity of data detection.
Therefore, it is desired to design a window which can null/suppress the power leakage and improve the effective channel sparsity.

\subsubsection{The Total Power of the Effective Channel Gain}
One subsequent question following the above derivations and discussions is whether there is a power change between the effective channel and the original channel response in the DD domain.
In this section, we derive the total power of the effective channel in the DD domain.
Since SFFT in \eqref{DDFilterIdealPulse} is an orthogonal transformation, it is straightforward that
\vspace{-2mm}
\begin{equation}
\sum_{k=0}^{N-1}\sum_{l=0}^{M-1} \left|w(k - {k_{{\nu _i}}}-\kappa_{\nu_i},l - {l_{{\tau _i}}})\right|^2 = \frac{1}{MN}\sum\limits_{n = 0}^{N - 1} \sum\limits_{m = 0}^{M - 1} \left| V\left[ {n,m} \right]U\left[ {n,m} \right] \right|^2, \forall i.\vspace{-2mm}
\end{equation}
Therefore, for normalized transmitter and receiver windows, i.e., $\sum\limits_{n = 0}^{N - 1} \sum\limits_{m = 0}^{M - 1} \left| V\left[ {n,m} \right]U\left[ {n,m} \right] \right|^2 = MN$, the total power of the effective channel in the DD domain is given by Equation \eqref{EffectiveChannelGain} on the top of next page.
\begin{figure*}[!t]
\vspace{-7mm}
\begin{align}
\sum_{k=0}^{N-1}\sum_{l=0}^{M-1} \left|{h_w}\left[ {k ,l } \right]\right|^2 &= \sum_{k=0}^{N-1}\sum_{l=0}^{M-1} \left|\sum_{i=1}^{P} {h_i} w(k-k_{\nu_i}-\kappa_{\nu_i}, l-l_{\tau_i}) {e^{ - j2\pi \frac{\left(k_{\nu_i} + \kappa_{\nu_i}\right)l_{\tau_i}}{NM} }}\right|^2 \notag\\[-1mm]
&\hspace{-35mm}= \sum_{k=0}^{N-1}\sum_{l=0}^{M-1} \left\{\sum_{i=1}^{P} \left|{h_i}\right|^2 \left|w(k-k_{\nu_i}-\kappa_{\nu_i}, l-l_{\tau_i})\right|^2 \right. \notag\\[-3mm]
&\hspace{-35mm}\left. + 2 \Re \underbrace{\left\{\sum_{i=1}^{P} \sum_{j\neq i}^{P} {h_i}{h^*_j} w(k\hspace{-1mm}-\hspace{-1mm}k_{\nu_i}\hspace{-1mm}-\hspace{-1mm}\kappa_{\nu_i}, l\hspace{-1mm}-\hspace{-1mm}l_{\tau_i})w^*(k\hspace{-1mm}-\hspace{-1mm}k_{\nu_j}\hspace{-1mm}-\hspace{-1mm}\kappa_{\nu_j}, l\hspace{-1mm}-\hspace{-1mm}l_{\tau_j}) {e^{ j2\pi \frac{\left(k_{\nu_i} + \kappa_{\nu_i}\right)l_{\tau_i} - \left(k_{\nu_j} + \kappa_{\nu_j}\right)l_{\tau_j}}{NM}}}\right\}}_{\mathrm{Channel\; inter-spread}}\right\}.\label{EffectiveChannelGain}
\end{align}
\vspace{-10mm}\par\noindent
	\hrulefill
	\vspace{-10mm}
\end{figure*}
We can observe that the second term in the bracket in \eqref{EffectiveChannelGain} is caused by the inter-spread between different paths with $\forall i \neq j$.
For the case of integer Doppler, i.e., $\kappa_{\nu_i} = 0$, $\forall i$, we have $\tau_i \neq \tau_j$ or $k_{\nu_i} \neq k_{\nu_j}$, $\forall i \neq j$.
As a result, a simple rectangular window yields $w(k-k_{\nu_i}, l-l_{\tau_i})w^*(k-k_{\nu_j}, l-l_{\tau_j}) = 0$, $\forall i\neq j$, $\forall k,l$, i.e., no inter-spread, and the total power of the effective channel in the DD domain is given by
\vspace{-2mm}
\begin{equation}
\sum_{k=0}^{N-1}\sum_{l=0}^{M-1} \left|{h_w}\left[ {k ,l } \right]\right|^2 = \sum_{i=1}^{P} \left|{h_i}\right|^2,\vspace{-2mm}
\end{equation}
which implies that the power is conserved in transforming the original channel response to the effective channel in the DD domain.
On the other hand, with the existence of fractional Doppler, when $\tau_i \neq \tau_j$, $\forall i \neq j$, we still have $w(k-k_{\nu_i}-\kappa_{\nu_i}, l-l_{\tau_i})w^*(k-k_{\nu_j}-\kappa_{\nu_j}, l-l_{\tau_j}) = 0$, $\forall i\neq j$, $\forall k,l$, for a rectangular window.
However, for two different paths experiencing the same delay but different Doppler shifts, i.e., $\tau_j = \tau_i$ and $k_{\nu_i} + \kappa_{\nu_i} \neq k_{\nu_j} + \kappa_{\nu_j}$, $\forall i \neq j$, their inter-spread is inevitable as $w(k-k_{\nu_i}-\kappa_{\nu_i}, l-l_{\tau_i})w^*(k-k_{\nu_j}-\kappa_{\nu_j}, l-l_{\tau_j}) \neq 0$, $\forall i\neq j$, $\forall k,l$.
{In this case, for a snapshot of the communication channel, the total power of the effective channel may change (either increase or decrease, depending on constructive or destructive channel inter-spread, respectively) compared to that of the original DD domain channel response.
In contrast, assuming independent channel coefficients for different paths, i.e., $E\left\{{h_i}{h^*_j}\right\} = 0$, $\forall i \neq j$, we have
\vspace{-2mm}
\begin{equation}
E\left\{\sum_{k=0}^{N-1}\sum_{l=0}^{M-1} \left|{h_w}\left[ {k ,l } \right]\right|^2\right\} = E\left\{\sum_{i=1}^{P} \left|{h_i}\right|^2\right\},\vspace{-2mm}
\end{equation}
which implies the average total power of the effective channel in the DD domain is identical to that of the original DD domain channel response.
In other words, although the mean of total power of effective channel in the DD domain preserves, the corresponding variance is increased due to the channel inter-spread.
However, the detection performance is dominated by the potential destructive channel inter-spread, which is not be favorable for reliable communications.
Thus, to reduce the potential channel inter-spread, a proper window design that can improve the effective channel sparsity is desirable.}

\vspace{-4mm}
\subsection{The Impact of Windowing on Effective Channel Estimation}
In this work, we adopt the channel estimation scheme proposed in \cite{RavitejaOTFSCE}, where a single pilot symbol is embedded in the DD domain and a guard space is inserted between the pilot symbol and data symbols.
In fact, to the best of our knowledge, the channel estimation scheme in \cite{RavitejaOTFSCE} is the first DD domain channel estimation method proposed for OTFS in the literature, which is simple and practical.
In this section, we investigate the impact of windowing on channel estimation performance based on the scheme in \cite{RavitejaOTFSCE}.
%
%
Let us assume that the only pilot symbol $x_p$ is inserted at the $\left[k_p,l_p\right]$-th DD grid and data symbols ${x_d}\left[ {k,l} \right]$ are arranged as follow\cite{RavitejaOTFSCE}
\vspace{-2mm}
\begin{equation}\label{PilotEmbedded}
x\left[ {k,l} \right] = \left\{ {\begin{array}{*{20}{c}}
	{{x_p}}&{k = {k_p},l = {l_p}},\\[-1mm]
		0&\begin{array}{l}
	{k_p} - 2{k_{\max }} - 2\hat k \le k \le {k_p} + 2{k_{\max }} + 2\hat k, k \neq {k_p},\\[-1mm]
	\hspace{22mm}{l_p} - {l_{\mathrm{max}} } \le l \le {l_p} + {l_{\mathrm{max}} }, l \neq {l_p},
	\end{array}\\[-1mm]
	{{x_d}\left[ {k,l} \right]}&{{\rm{otherwise}}},
	\end{array}} \right.\vspace{-2mm}
\end{equation}
where $k_{\max } \in \mathbb{Z}^+$ and $l_{\max } \in \mathbb{Z}^+$ denote the maximum Doppler and delay indices, respectively, $\hat k \in \mathbb{Z}^+$ denotes the additional guard to mitigate the spread due to fractional Doppler and $\hat k \in \left\{0,\ldots,  \lfloor\frac{N-4k_{\max }-1}{4}\rfloor \right\}$.
Increasing $\hat k$ would potentially increase the channel estimation performance while reduces the spectral efficiency, as the signaling overhead increases with $\hat k$, i.e., the total signaling overhead is $\left(2{l_{\mathrm{max}} } +1\right)\left(4{k_{\max }} + 4\hat k + 1\right)$.
%

The estimation of the effective channel is based on the received signals in the DD domain, which are given by
\vspace{-3mm}
\begin{equation}\label{EEChannelEstimation}
y\left[k,l\right] = x_p {h_w}\left[ {\left(k-k_p\right)_N ,\left(l-l_p\right)_M } \right] + I\left[k,l\right] + \sum\limits_{k' = 0}^{N - 1} {\sum\limits_{l' = 0}^{M - 1} {z\left[ {k',l'} \right]} } {v_z}\left[ {k - k',l - l'} \right],\vspace{-2mm}
\end{equation}
where ${k_p} - {k_{\max }} - \hat k \le k \le {k_p} + {k_{\max }} + \hat k$ and ${l_p} \le l \le {l_p} + {l_{\mathrm{max}} }$.
According to \cite{RavitejaOTFSCE}, the effective DD domain channel can be estimated by
\vspace{-2mm}
\begin{equation}\label{DDCEFormula}
	{\hat h_w}\left[ {\left(k-k_p\right)_N ,\left(l-l_p\right)_M } \right] = \frac{y\left[k,l\right]}{x_p}, \text{if} \left|y\left[k,l\right]\right| \ge 3\sqrt{N_0}.\vspace{-2mm}
\end{equation}
In \eqref{EEChannelEstimation}, $I\left[k,l\right]$ denotes the interference spread from the data symbols due to the existence of fractional Doppler and it is given by
\vspace{-2mm}
\begin{equation}\label{InterferenceCE}
I\left[k,l\right] = \sum_{k' \notin \left[{k_p} - 2{k_{\max }} - 2\hat k , {k_p} + 2{k_{\max }} + 2\hat k\right]} \sum_{l'=0}^{l_{\mathrm{max}}} x\left[ {k',\left(l-l'\right)_M} \right] {h_w}\left[ {\left(k-k'\right)_N ,l' } \right].\vspace{-2mm}
\end{equation}
We can observe that in the delay domain, only $l_{\mathrm{max}}+1$ symbols before $l$ affect the received symbol on $l$.
On the other hand, in the Doppler domain, all the data symbols outside the guard space $k' \notin \mathcal{K} = \left[{k_p} - 2{k_{\max }} - 2\hat k , {k_p} + 2{k_{\max }} + 2\hat k\right]$ affect the received symbol on $k$.
Due to the existence of the interference term $I\left[k,l\right]$, the channel estimation in \eqref{DDCEFormula} suffers from an error floor even increasing the system signal-to-noise ratio (SNR).
Furthermore, the effective channel estimation error deteriorates the detection performance and results in an error floor in data detection performance.
Note that when using the full guard space\cite{RavitejaOTFSCE}, i.e., $4{k_{\max }} + 4\hat k + 1 = N$, the interference term in \eqref{InterferenceCE} would disappear and there is no error floor in the effective channel estimation.
However, it requires a higher signaling overhead of $\left(2{l_{\mathrm{max}} } +1\right)N$ compared to that of the scheme in \eqref{PilotEmbedded}.

In the following, we derive the interference power to investigate the impact of windowing on the effective channel estimation performance.
Since the transmitted data symbols are independent, the interference power can be calculated as \eqref{InterferenceLevel} {on the top of this page,}
\begin{figure*}[!t]
	\vspace{-8mm}
	\begin{align}\label{InterferenceLevel}
	&E\left\{ {{{\left| {I\left[ {k,l} \right]} \right|}^2}} \right\} 
	= \sum\limits_{k' \notin \mathcal{K}} {\sum\limits_{l' = l-l_{\mathrm{max}}}^{{l }} {E\left\{ {{{\left| {x\left[ {k',l'} \right]} \right|}^2}} \right\} E\left\{ {{{\left| {{h_w}\left[ {{{\left( {k - k'} \right)}_N},{{\left( {l - l'} \right)}_M}} \right]} \right|}^2}} \right\}} }\notag\\[-1mm]
	&= \sum\limits_{k' \notin \mathcal{K}} \sum\limits_{l' = l-l_{\mathrm{max}}}^{{l }} \hspace{-1mm} E\left\{ {{{\left| {x\left[ {k',l'} \right]} \right|}^2}} \right\} {E\left\{ {{{\left| {\sum\limits_{i = 1}^P {{h_i}} w\left( {{{\left( {k \hspace{-1mm}-\hspace{-1mm} k'} \right)}_N} \hspace{-1mm}-\hspace{-1mm} {k_{{\nu _i}}} \hspace{-1mm}-\hspace{-1mm} {\kappa _{{\nu _i}}},{{\left( {l \hspace{-1mm}-\hspace{-1mm} l'} \right)}_M} \hspace{-1mm}-\hspace{-1mm}l_{\tau_i}} \right){e^{ - j2\pi \frac{{\left( {{k_{{\nu _i}}} + {\kappa _{{\nu _i}}}} \right){l_{{\tau _i}}}}}{{NM}}}}} \right|}^2}} \right\}}  \notag\\[-1mm]
	& \mathop  = \limits^{(a)}  \sum\limits_{k' \notin \mathcal{K}} \sum\limits_{ l-l' \in [0,l_{\mathrm{max}}], {{\left( {l - l'} \right)}_M} = l_{\tau_i}} {{E\left\{ {{{\left| {\sum\limits_{i = 1}^P {{h_i}} w\left( {{{\left( {k - k'} \right)}_N} - {k_{{\nu _i}}} - {\kappa _{{\nu _i}}},0} \right){e^{ - j2\pi \frac{{\left( {{k_{{\nu _i}}} + {\kappa _{{\nu _i}}}} \right){l_{{\tau _i}}}}}{{NM}}}}} \right|}^2}} \right\}} }.
	\end{align}
	\vspace{-10mm}\par\noindent
	\hrulefill
	\vspace{-10mm}
\end{figure*}
where the equality $(a)$ is obtained since only the data symbols on ${{\left( {l - l'} \right)}_M} = l_{\tau_i}$ in the summation over $l'$ affect the received symbol on $l$ with adopting a rectangular window in the delay domain, i.e., $V\left[ {n,m} \right] = V\left[ {n,m'} \right] = U\left[ {n,m} \right] = U\left[ {n,m'} \right]$, $\forall n,m,m'$, and $E\left\{ {{{\left| {x\left[ {k',l'} \right]} \right|}^2}} \right\} = 1$.
Assuming independent channel coefficients, i.e., $E\left\{{h_i}{h^*_j}\right\} = 0$, $\forall i \neq j$, \eqref{InterferenceLevel} becomes
\vspace{-2mm}
\begin{equation}
E\left\{ {{{\left| {I\left[ {k,l} \right]} \right|}^2}} \right\}  = \sum\limits_{k' \notin \mathcal{K}} {\sum\limits_{i = 1}^P {E\left\{ {{{\left| {{h_i}} \right|}^2}} \right\}} {{\left| {w\left( {{{\left( {k - k'} \right)}_N} - {k_{{\nu _i}}} - {\kappa _{{\nu _i}}},0} \right)} \right|}^2}}.\vspace{-2mm}
\end{equation}

It can be observed that the interference power is determined by the window response at ${{\left( {k - k'} \right)}_N} - {k_{{\nu _i}}} - {\kappa _{{\nu _i}}}$.
Thanks to the guard space, the window response $\left| {w\left( {{{\left( {k - k'} \right)}_N} - {k_{{\nu _i}}} - {\kappa _{{\nu _i}}},0} \right)} \right|$ lies in its sidelobe and becomes almost a constant, as shown in Fig. \ref{FractionalIDI}.
Assume $\left| {w\left( {{{\left( {k - k'} \right)}_N} - {k_{{\nu _i}}}- {\kappa _{{\nu _i}}},0} \right)} \right| \approx \mathrm{SL}_{w}$ for ${k' \notin \left[ {{k_p} - 2{k_{\max }} - 2\hat k,{k_p} + 2{k_{\max }} + 2\hat k} \right]}$ and $k \in \left[{k_p} - {k_{\max }} - \hat k , {k_p} + {k_{\max }} + \hat k\right]$.
Considering a normalized channel power gain, i.e., $\sum\limits_{i = 1}^P {E\left\{ {{{\left| {{h_i}} \right|}^2}} \right\}} = 1$, the average interference power can be approximated by
\vspace{-2mm}
\begin{equation}
E\left\{ {{{\left| {I\left[ {k,l} \right]} \right|}^2}} \right\}  \approx \left(N-4{k_{\max }} - 4\hat k - 1\right) \mathrm{SL}_{w}^2,\vspace{-2mm}
\end{equation}
where $\mathrm{SL}_{w}$ denotes the sidelobe level of the adopted window.
For instance, as shown in Fig. \ref{FractionalIDI}, we have $\mathrm{SL}_{w} \approx \frac{1}{N}$ for the case of rectangular window.
According to \eqref{EEChannelEstimation}, in high SNR regime, i.e., $N_0 \to 0$, the MSE of the effective channel estimation in the guard space is given by
\vspace{-2mm}
\begin{align}\label{ChannelESTIMATIONMSE}
	\mathrm{MSE} &= \sum_{k = {k_p} - {k_{\max }} - \hat k}^{{k_p} + {k_{\max }} + \hat k} \sum_{l = {l_p}}^{{l_p} + {l_{\mathrm{max}} }}E\left\{ {{{\left| {h_w}\left[ {\left(k-k_p\right)_N ,\left(l-l_p\right)_M } \right] - {\hat{h}_w}\left[ {\left(k-k_p\right)_N ,\left(l-l_p\right)_M } \right] \right|}^2}} \right\} \notag\\[-1mm]
	&= \sum_{k = {k_p} - {k_{\max }} - \hat k}^{{k_p} + {k_{\max }} + \hat k} \sum_{l = {l_p}}^{{l_p} + {l_{\mathrm{max}} }} \frac{E\left\{ {{{\left| {I\left[ {k,l} \right]} \right|}^2}} \right\}}{\left|x_p\right|^2} \notag\\[-1mm]
	&\approx \left(N-4{k_{\max }} - 4\hat k - 1\right) \left(2{k_{\max }} + 2\hat k+1\right) \left(l_{\mathrm{max}} + 1\right)\mathrm{SL}_{w}^2.
\end{align}
\vspace{-8mm}\par\noindent


Now, we can observe that in the high SNR regime, the error floor level in the effective channel estimation in \eqref{ChannelESTIMATIONMSE} depends on the additional guard $\hat k$ and the sidelobe level $\mathrm{SL}_{w}$ of the designed window response.
{It can be seen that the MSE of the effective channel estimation is a quadratic function with respect to $\hat k$.
After some mathematical manipulations, it can be seen that when $\frac{N-8{k_{\max }-3}}{4} \le 0$, increasing $\hat k $ in the range of $\left\{0,\ldots,  \lfloor\frac{N-4k_{\max }-1}{4}\rfloor \right\}$ always results in a lower error floor level at the expense of more signaling overhead.
On the other hand, when $\frac{N-8{k_{\max }-3}}{4} \ge 1$, increasing $\hat k $ first increases and then decreases the MSE of effective channel estimation.
This is because for large $N$, increasing $\hat k $ introduces more entries to be estimated within the guard space, thereby might increasing channel estimation error.
Further increasing $\hat k $ reduces the IDI caused by the data symbols and thus reduces the effective channel estimation MSE, but it also consumes more signaling overhead.}
More importantly, as shown in \eqref{ChannelESTIMATIONMSE}, a proper design of window response can achieve a low sidelobe level at the first place, which can effectively decrease the error floor level with a relatively small $\hat k$.
In fact, a window response with a low sidelobe level can enhance the effective channel sparsity, which can improve the channel estimation performance.
Moreover, as the TX and RX windows have the same impact on the effective channel in the DD domain in \eqref{DDIOChannelDiscreteIdealPulseFractional}, imposing a window at either the transmitter or the receiver will result in the same channel estimation error floor.


\vspace{-4mm}
\subsection{The Impact of TX Window on Average Transmit Power}
In this section, we investigate the impact of TX window on the total average transmit power of OTFS system.
Following the vectorization form in \eqref{VectorForm}, the transmitted signal in the time domain can be written as
\vspace{-4mm}
\begin{equation}
	\mathbf{s}_t = \left(\mathbf{I}_{N} \otimes  \mathbf{F}_{M}^{\mathrm{H}}  \right) \mathbf{U} \left( \mathbf{F}_{N}^{\mathrm{H}} \otimes \mathbf{F}_{M}\right) \mathbf{x}_{\mathrm{DD}},\vspace{-2mm}
\end{equation}
and the average transmit power of OTFS modulator is
\vspace{-2mm}
\begin{align}
E\left\{\mathbf{s}^{\mathrm{H}}_t \mathbf{s}_t\right\} &= \Tr\left\{E\left\{ \mathbf{s}_t \mathbf{s}^{\mathrm{H}}_t\right\}\right\} \notag\\[-1mm]
   &= \Tr\left\{E\left\{ \left(\mathbf{I}_{N} \otimes  \mathbf{F}_{M}^{\mathrm{H}} \right) \mathbf{U} \left(\mathbf{F}_{M} \otimes \mathbf{F}_{N}^{\mathrm{H}}  \right) \mathbf{x}_{\mathrm{DD}} \mathbf{x}^{\mathrm{H}}_{\mathrm{DD}} \left(\mathbf{F}_{N} \otimes \mathbf{F}^{\mathrm{H}}_{M} \right) \mathbf{U}^{\mathrm{H}} \left(\mathbf{I}_{N} \otimes \mathbf{F}_{M} \right) \right\}\right\} \notag\\[-1mm]
   &= \Tr\left\{\left( \mathbf{I}_{N} \otimes  \mathbf{F}_{M}^{\mathrm{H}}\right) \mathbf{U} \mathbf{U}^{\mathrm{H}} \left(\mathbf{I}_{N} \otimes  \mathbf{F}_{M}\right)\right\}= \Tr\left\{ \mathbf{U} \mathbf{U}^{\mathrm{H}} \right\}= \sum_{m=0}^{M-1}\sum_{n=0}^{N-1} \left|U{\left[m,n\right]}\right|^2.
\end{align}
\vspace{-6mm}\par\noindent
We can observe that the average power of the time domain transmitted signal is determined by the summation of square of the TX window in the TF domain.
In fact, the TX window $\left|U{\left[m,n\right]}\right|^2$ can be interpreted as the power allocation for the $\left(nM+m\right)$-th symbol in one OTFS frame in the TF domain.

\vspace{-4mm}
\subsection{The Impact of RX Window on Noise Covariance Matrix}
Due to the existence of RX window, the noise covariance matrix at the OTFS demodulator is given by
\vspace{-4mm}
\begin{align}
\mathbf{C}_{\mathbf{z}} &= E\left\{\left(\mathbf{F}_{N} \otimes \mathbf{F}_{M}^{\mathrm{H}} \right)\mathbf{V} \left(\mathbf{I}_N \otimes \mathbf{F}_{M} \right) \mathbf{w} \mathbf{w}^{\mathrm{H}} \left(\mathbf{I}_N \otimes \mathbf{F}^{\mathrm{H}}_{M} \right) \mathbf{V}^{\mathrm{H}}\left(\mathbf{F}^{\mathrm{H}}_{N} \otimes \mathbf{F}_{M} \right) \right\} \notag\\[-1mm]
&=N_0\left(\mathbf{F}_{N} \otimes \mathbf{F}_{M}^{\mathrm{H}} \right)\mathbf{V} \mathbf{V}^{\mathrm{H}} \left(\mathbf{F}^{\mathrm{H}}_{N} \otimes \mathbf{F}_{M}\right). \label{NoiseConvarianceMatrix}
\end{align}
\vspace{-8mm}\par\noindent

When $\mathbf{V} \mathbf{V}^{\mathrm{H}} = \mathbf{I}_{MN}$, we have $\mathbf{C}_{\mathbf{z}} = N_0 \mathbf{I}_{MN}$, implying that the effective noise in the DD domain is still white even with a RX window.
Note that $\mathbf{V} \mathbf{V}^{\mathrm{H}} = \mathbf{I}_{MN}$ implies that we can only impose a constant-modulus RX window in the TF domain, i.e., $\left|V_{m,n}\right| = 1$.
When $\mathbf{V} \mathbf{V}^{\mathrm{H}} \neq \mathbf{I}_{MN}$, the covariance matrix of the effective noise in the DD domain is not an identity matrix.
It indicates that using a RX window $\mathbf{V} \mathbf{V}^{\mathrm{H}} \neq \mathbf{I}_{MN}$ causes a colored noise, which imposes a challenge for the data detection\footnote{Note that although a whitening filter can be applied in the OTFS demodulator to whiten the colored noise, the whitening filter would completely or partially reverse the effect of the RX window, resulting a non-sparse effective channel in the DD domain degrading the channel estimation performance.}.

\vspace{-4mm}
\section{Window Designs for OTFS}
In this section, we design the TF domain window based on the above discussed properties of OTFS systems.
In particular, when CSI is available at both the transmitter and receiver, we propose an optimal TX window design to minimize the data detection MSE.
On the other hand, when CSI is not known at the transmitter but can be estimated at the receiver, we first discuss the ideal window response, which provides guidelines for practical window designs.
Based on that, we propose to apply the DC window to enhance the effective channel sparsity, which will significantly improve the performance of both channel estimation and data detection.

\vspace{-4mm}
\subsection{With CSI at OTFS Transceiver}
{As mentioned before, the CSI\footnote{Here, the CSI denotes the information in either the TF domain or the DD domain as they are interchangeable as shown in \eqref{TFDomainChannel} and \eqref{DDDomainChannel}.} can be estimated at the OTFS receiver (CSIR) with adopting the channel estimation scheme in \eqref{DDCEFormula}.
On the other hand, due to the channel stability in the DD domain\cite{Hadani2017orthogonal}, acquiring CSI is possible at the transmitter (CSIT) via channel tracking \cite{SimeoneChannelTracking} or user feedback \cite{WenChaoKai}.
As the CSI is available, window design should focus on improving the detection performance rather than enhancing the channel sparsity.}
In particular, the minimum mean squared error (MMSE) detector is adopted to facilitate the window design due to its tractability and promising performance\footnote{Note that some tailor-made OTFS detectors, e.g. sum-product algorithm (SPA)-based detector, may be more efficient as its computational complexity depends on the number of non-zero entries in the effective DD domain channel.
	However, analyzing the performance of such detectors depends on the specific implementation details and is generally intractable.
	Therefore, we adopt a commonly adopted MMSE detector to facilitate the window design.
	The simulation results of an OTFS system with an SPA detector will be evaluated in Section V.}.
In particular, we first derive the detection MSE as a function of TX/RX window and then design the TX window to minimize the detection MSE.

According to \eqref{VectorForm}, the MMSE estimation of ${{{\bf{ x}}}_{{\rm{DD}}}}$ is given by
\vspace{-2mm}
\begin{equation}\label{MMSE}
{{{\bf{\hat x}}}_{{\rm{DD}}}} = {\bf{H}}_{{\rm{DD}}}^{\rm{H}}{\left( {{\bf{H}}_{{\rm{DD}}}^{}{\bf{H}}_{{\rm{DD}}}^{\rm{H}} + {\bf{C}}_{\bf{z}}^{}} \right)^{ - 1}}{{\bf{y}}_{{\rm{DD}}}},\vspace{-2mm}
\end{equation}
and the corresponding estimation error covariance matrix in the DD domain is
\vspace{-2mm}
\begin{align}\label{MMSERxTxWINDOW}
{{\bf{E}}_{{\rm{DD}}}} &= {\left( {{{\bf{I}}_{MN}} + {\bf{H}}_{{\rm{DD}}}^{\rm{H}}{\bf{C}}_{\bf{z}}^{ - 1}{\bf{H}}_{{\rm{DD}}}^{}} \right)^{ - 1}},\notag\\[-1mm]
& = {\left( {{{\bf{I}}_{MN}} + \frac{1}{N_0}\left(\mathbf{F}_{N} \otimes \mathbf{F}_{M}^{\mathrm{H}} \right) \mathbf{U}^{\mathrm{H}} \left( \mathbf{I}_{N} \otimes \mathbf{F}_{M} \right){{{{{\bf{H}}}^{\mathrm{H}}_t}}}{{{{{\bf{H}}}^{}_t}}} \left( \mathbf{I}_{N} \otimes \mathbf{F}^{\mathrm{H}}_{M} \right) \mathbf{U} \left( \mathbf{F}^{\mathrm{H}}_{N} \otimes \mathbf{F}_{M} \right)} \right)^{ - 1}},
\end{align}
\vspace{-8mm}\par\noindent
where the average MSE per symbol in the DD domain is obtained by
\vspace{-1mm}
\begin{equation}
\sigma^2_e = \frac{\mathrm{Tr} \{{{\bf{E}}_{{\rm{DD}}}}\}}{MN}.\vspace{-2mm}
\end{equation}
We can observe that the RX window is irrelevant to the detection MSE, while the TX window may change the detection performance.
This is because the RX window can be viewed as a preprocessing on the received signal, which affects the effective channel and noise covariance matrix in the DD domain at the same time, as shown in \eqref{VectorForm}, thereby will not change the detection MSE.
In the following, we optimize the TX window to minimize the detection MSE.

Considering the ideal pulse shaping filter, the time domain channel can be decomposed as ${{{{{\bf{H}}}_t}}} = {\left( \mathbf{I}_{N} \otimes \mathbf{F}^{\mathrm{H}}_{M} \right) {{\widetilde {\bf{H}}}_{{\rm{TF}}}} \left( \mathbf{I}_{N} \otimes \mathbf{F}_{M} \right)}$, where ${{\widetilde {\bf{H}}}_{{\rm{TF}}}}$ is a diagonal matrix denoting the TF domain channel matrix.
Therefore, the semi-positive definite matrix $\frac{1}{N_0}{{{{{\bf{H}}}^{\mathrm{H}}_t}}}{{{{{\bf{H}}}^{}_t}}} $ can be decomposed as $\frac{1}{N_0}{{{{{\bf{H}}}^{\mathrm{H}}_t}}}{{{{{\bf{H}}}^{}_t}}} = \left( \mathbf{I}_{N} \otimes \mathbf{F}^{\mathrm{H}}_{M} \right)\mathbf{\Lambda}\left( \mathbf{I}_{N} \otimes \mathbf{F}_{M} \right) $, where $\mathbf{\Lambda} = \frac{1}{N_0}{{\widetilde {\bf{H}}}_{{\rm{TF}}}}^{\mathrm{H}} {{\widetilde {\bf{H}}}_{{\rm{TF}}}} =  \diag\{\lambda_{1,1},\lambda_{1,2},\ldots,\lambda_{m,n},\ldots,\lambda_{M,N}\}$ collects the non-negative real-valued eigenvalues.
Now, the estimation error covariance matrix in \eqref{MMSERxTxWINDOW} can be rewritten as 
\vspace{-2mm}
\begin{equation}
	{{\bf{E}}_{{\rm{DD}}}} = {\left( {{{\bf{I}}_{MN}} + \left(\mathbf{F}_{N} \otimes \mathbf{F}_{M}^{\mathrm{H}} \right) \mathbf{U}^{\mathrm{H}} \mathbf{\Lambda} \mathbf{U} \left( \mathbf{F}^{\mathrm{H}}_{N} \otimes \mathbf{F}_{M} \right)} \right)^{ - 1}}.\vspace{-2mm}
\end{equation}
Set $\bf{Q} = \left(\mathbf{F}_{N} \otimes \mathbf{F}_{M}^{\mathrm{H}} \right) \mathbf{U}^{\mathrm{H}} $, the trace of the estimation error covariance matrix is
\vspace{-2mm}
\begin{align}
	\mathrm{Tr} \{{{\bf{E}}_{{\rm{DD}}}}\} &= \mathrm{Tr}\left\{\left( {{{\bf{I}}_{MN}} + \bf{Q}\mathbf{\Lambda}\bf{Q}^{\mathrm{H}}} \right)^{ - 1}\right\} = \mathrm{Tr}\left\{ {{{\mathbf{I}}_{MN}} - \bf{Q}\left(\mathbf{\Lambda}^{-1} + \bf{Q}^{\mathrm{H}}\bf{Q} \right)^{-1}\bf{Q}^{\mathrm{H}}}\right\} \notag\\[-1mm]
	& = MN-\mathrm{Tr}\left\{ { \left(\mathbf{\Lambda}^{-1} + \bf{Q}^{\mathrm{H}}\bf{Q} \right)^{-1}\bf{Q}^{\mathrm{H}}\bf{Q}}\right\} = MN-\mathrm{Tr}\left\{ \mathbf{I}_{MN} - \left(\mathbf{\Lambda}^{-1} + \bf{Q}^{\mathrm{H}}\bf{Q} \right)^{-1} \mathbf{\Lambda}^{-1} \right\} \notag\\[-1mm]
	& = \mathrm{Tr}\left\{\left(\mathbf{I}_{MN} + \mathbf{U} \mathbf{U}^{\mathrm{H}} \mathbf{\Lambda} \right)^{-1}  \right\} = \sum_{m=1}^{M} \sum_{n=1}^{N} \frac{1}{\lambda_{m,n} \left|U{\left[m,n\right]}\right|^2 + 1}.\label{DetectionMSE}
\end{align}
\vspace{-7mm}\par\noindent

%
To minimize the detection MSE, we define $x_{m,n} = \left|U{\left[m,n\right]}\right|^2$ and formulate the following optimization problem:
\vspace{-4mm}
\begin{align}\label{TxWindowOptmization}
\underset{{x}_{m,n}}{\mino}\,\, &\frac{1}{MN} \sum_{m=1}^{M} \sum_{n=1}^{N} \frac{1}{\lambda_{m,n} x_{m,n} + 1}  \\[-1mm]
\notag\mbox{s.t.}\;\;
&\mbox{{C1}}:\; x_{m,n} \ge 0,\forall m,n, \notag\\[-1mm]
&\mbox{{C2}}:\; \frac{1}{MN} \sum_{m=1}^{M} \sum_{n=1}^{N} x_{m,n} \le 1,\notag
\end{align}
\vspace{-8mm}\par\noindent
where constraint C2 is imposed to guarantee a normalized TX window, i.e., a fixed average transmit power.

\begin{figure}[t]
	\centering\vspace{-5mm}
	\includegraphics[width=3.8in]{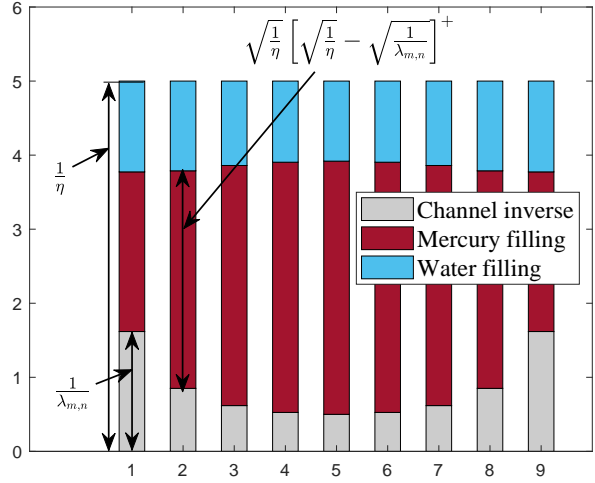}\vspace{-7mm}
	\caption{An illustration of mercury/water filling in \eqref{OptimalTxWindowMercury}.}\vspace{-10mm}
	\label{MercuryWaterFilling}%
\end{figure}

The formulated problem in \eqref{TxWindowOptmization} is a standard convex optimization problem and we can obtain its optimal solution via solving the corresponding dual problem.
The optimal solution of \eqref{TxWindowOptmization} for the TX window $U{\left[m,n\right]}$ is given by
\vspace{-2mm}
\begin{equation}\label{OptimalTxWindow}
x_{m,n} = \left|U{\left[m,n\right]}\right|^2 = \left[\sqrt{\frac{1}{\eta \lambda_{m,n}}} - \frac{1}{\lambda_{m,n}}\right]^+,\vspace{-2mm}
\end{equation}
where $\eta$ is a dual variable which can be obtained via exhaustive search such that $\frac{1}{MN} \sum\limits_{m=1}^{M} \sum\limits_{n=1}^{N} x_{m,n} = 1$.
To obtain $U{\left[m,n\right]}$, we adopt a real-valued TX window $U{\left[m,n\right]} = \sqrt{x_{m,n}}$ since the derived detection MSE in \eqref{DetectionMSE} does not depend on the phase of $U{\left[m,n\right]}$.
We can observe that the optimal TX window in \eqref{OptimalTxWindow} is similar to but different from the conventional water-filling power allocation scheme\cite{Tse2005}.
To retain some intuition, we follow the interpretation of the mercury/water filling in \cite{LozanoMercuryWaterFilling} and rewrite \eqref{OptimalTxWindow} as follows:
\vspace{-2mm}
\begin{equation}\label{OptimalTxWindowMercury}
x_{m,n} = \left[\frac{1}{\eta} - \underbrace{\sqrt{\frac{1}{\eta}} \left[\sqrt{\frac{1}{\eta}} - \sqrt{\frac{1}{\lambda_{m,n}}}\right]^+}_{\mathrm{Mercury\;filling}} -\frac{1}{\lambda_{m,n}}\right]^+,\vspace{-2mm}
\end{equation}
where the second term $\sqrt{\frac{1}{\eta}} \left[\sqrt{\frac{1}{\eta}} - \sqrt{\frac{1}{\lambda_{m,n}}}\right]^+$ denotes the mercury filling and $x_{m,n}$ is the water filling.
To visualize the optimal TX window, we show the process of mercury/water filling for a case of $MN = 9$ in Fig. \ref{MercuryWaterFilling}.
{The inverse of each eigen-channel $\frac{1}{\lambda_{m,n}}$ is firstly filled in each vessel.
Then, the mercury with a height of $\sqrt{\frac{1}{\eta}} \left[\sqrt{\frac{1}{\eta}} - \sqrt{\frac{1}{\lambda_{m,n}}}\right]^+$ is poured into each vessel.
Finally, water is filled into all vessels until reaching the water level  $\frac{1}{\eta}$.
The depth of water is the power allocated to the corresponding eigen-channel.
It can be seen that the higher the channel power gain ${\lambda_{m,n}}$, the higher the mercury level, and vice versa.
The resulting heights in all vessels after filling the mercury are balanced compared to the channel inverse in the TF domain $\frac{1}{\lambda_{m,n}}$.
In fact, the mercury filling can be interpreted as a pre-equalization against the doubly selective fading in the TF domain, leading to a channel not so time-variant and frequency-selective.
Besides, we can observe that when $\eta \ge \lambda_{m,n}$, neither mercury nor water should be filled into the vessel, i.e., the condition of the corresponding eigen-channel is unsatisfactory and should not be exploited for signal transmission.}


\vspace{-4mm}
\subsection{Without CSI at OTFS Transmitter}
For a more practical case where CSI is not available at the transceiver but can be estimated at the receiver, a universal window design to enhance the effective channel sparsity is preferred and beneficial to improve the channel estimation performance, which will be presented as follows.
%

\subsubsection{Ideal Window}
To facilitate the window design, we consider a separable TF domain window as follows:
\vspace{-2mm}
\begin{equation}
V\left[ {n,m} \right] = V_{\nu}\left[ {n} \right] V_{\tau}\left[ {m} \right]\;\text{and}\;
U\left[ {n,m} \right] = U_{\nu}\left[ {n} \right] U_{\tau}\left[ {m} \right],\vspace{-2mm}
\end{equation}
where $V_{\nu}\left[ {n} \right]$ and $U_{\nu}\left[ {n} \right]$ denote the RX and TX windows in the Doppler domain, respectively, and $V_{\tau}\left[ {m} \right]$ and $U_{\tau}\left[ {m} \right]$ denote the RX and TX windows in the delay domain, respectively.
As a result, the window response in the DD domain in \eqref{DDFilterIdealPulse} can be decomposed as
\vspace{-2mm}
\begin{equation}\label{ArbitraryWindow}
w(k \hspace{-1mm} - \hspace{-1mm}{k_{{\nu _i}}}\hspace{-1mm}-\hspace{-1mm}\kappa_{\nu_i},l \hspace{-1mm}-\hspace{-1mm} {l_{{\tau _i}}}) = \mathcal{G}_N\left({k - {k_{{\nu _i}}}  -\kappa_{\nu_i}}\right) \mathcal{F}_M \left({l - {l_{{\tau _i}}}}\right),\vspace{-2mm}
\end{equation}
where 
\vspace{-2mm}
\begin{align}
	\mathcal{G}_N\left({k - {k_{{\nu _i}}}  -\kappa_{\nu_i}}\right) &= \frac{1}{N}\sum\limits_{n = 0}^{N - 1}
	V_{\nu}\left[ {n} \right]U_{\nu}\left[ {n} \right]{e^{ - j2\pi n\frac{\left( {k-k_{\nu_i} - \kappa_{\nu_i}} \right)}{N}}}\;\text{and}\label{WindowResponseFunctionDoppler}\\[-1mm]
	\mathcal{F}_M \left({l - {l_{{\tau _i}}}}\right) &= \frac{1}{M}\sum\limits_{m = 0}^{M - 1} V_{\tau}\left[ {m} \right]U_{\tau}\left[ {m} \right]{e^{j2\pi m \frac{\left( l-l_{\tau_i} \right)}{M}}}.
\end{align}
\vspace{-8mm}\par\noindent

Combining \eqref{DDIOChannelDiscreteIdealPulseFractional} and \eqref{ArbitraryWindow}, the effective channel in the DD domain can be rewritten as
\vspace{-2mm}
\begin{equation}
{h_w}\left[ {k ,l } \right]
=\sum\limits_{i = 1}^P {h_i} \mathcal{G}_N\left({k - {k_{{\nu _i}}}  -\kappa_{\nu_i}}\right) \mathcal{F}_M \left({l - {l_{{\tau _i}}}}\right) {e^{ - j2\pi \frac{\left(k_{\nu_i} + \kappa_{\nu_i}\right)l_{\tau_i}}{NM} }}.\vspace{-2mm}\label{OTFS_DeModIdealPulseIIIFractional}
\end{equation}
Since the delay resolution is usually sufficient and there is only negligible channel spread in the delay domain \cite{RavitejaOTFS}, the optimal window in the delay domain should be maintained as the rectangular window, i.e., $\mathcal{F}^{\mathrm{Ideal}}_M \left({l}\right) = \mathcal{F}^{\mathrm{Rect}}_M \left({l }\right)$.
Besides, with the existence of the fractional Doppler, as $-\frac{1}{2} < \kappa_{\nu_i} < \frac{1}{2}$, the ideal window in the Doppler domain is given by:
\vspace{-2mm}
\begin{equation}\label{IdealWindow}
{{\cal G}^{\mathrm{Ideal}}_N}\left( k \right) = \left\{ {\begin{array}{*{20}{c}}
	{1,}&{ - 0.5 \le k \le 0.5,}\\[-1mm]
	{0,}&{{\rm{otherwise.}}}
	\end{array}} \right.\vspace{-2mm}
\end{equation}
In Fig. \ref{DifferentWindow}, we illustrate that the ideal window can tolerate the fractional Doppler shift without sacrificing any channel gain and causing any channel spread.
However, to implement such an ideal window response, an infinite length of window in the time domain is needed, i.e., $N \to \infty$.
Recall that the fractional Doppler is caused by the finite $N$.
Therefore, the ideal window in \eqref{IdealWindow} is not realizable in practice.

\subsubsection{Dolph-Chebyshev Window}
In what follows, we propose to apply Dolph-Chebyshev (DC) window at the transmitter or the receiver to improve the channel sparsity when CSI is not available.
In fact, it has been proved that the DC window is effective \cite{Dolph1946} in the sense that: 1) given the specified sidelobe level, the width of the mainlobe in the window response is the narrowest; or 2) given the fixed mainlobe width, the sidelobe level is minimized.
Note that the effective channel only has a considerably large entry when it is located in the mainlobe of the window response function in \eqref{WindowResponseFunctionDoppler}.
Therefore, given any channel sparsity requirement, the channel spread to other Doppler indices is reduced to the largest degree by using a DC window.

\begin{figure}[t]
	\centering\vspace{-5mm}
	\includegraphics[width=3.8in]{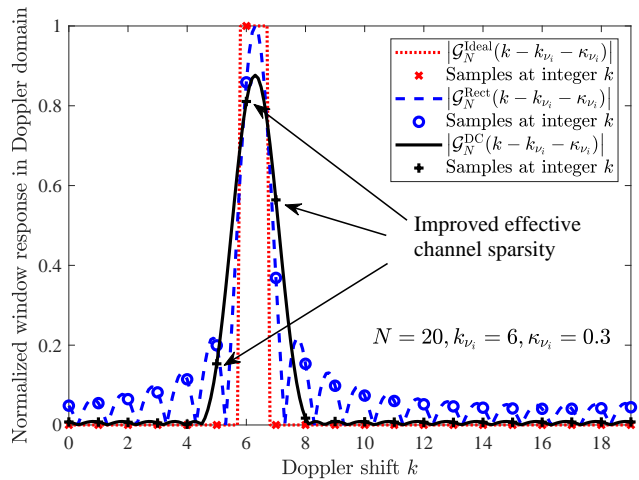}\vspace{-7mm}
	\caption{The effective channel in the Doppler domain with different types of windows and fractional Dopplers.}\vspace{-5mm}
	\label{DifferentWindow}%
\end{figure}

Particularly, if the required mainlobe width of the window response in the Doppler domain is $k_{\mathrm{main}} > 1$, the number of non-negligible effective channel spreading of each path in the Doppler domain is no more than $k_{\mathrm{main}}$, i.e., the effective channel sparsity is improved.
In this case, the lowest sidelobe level achieved by the DC window is \cite{WeiBeamWidthControl}
\vspace{-2mm}
\begin{equation}
\mathrm{SL}_w [\mathrm{dB}] = -20 \log_{10} {{\cosh \left( {\frac{N}{2}{{\cosh }^{-1}}\left( \frac{3-\cos\left(\frac{k_{\mathrm{main}}}{2}\right)}{1+\cos\left(\frac{k_{\mathrm{main}}}{2}\right)} \right)} \right)}}.\vspace{-2mm}
\end{equation}
In this paper, to reveal the insights of employing TX/RX windows, we only adopt the DC window at the transmitter or the receiver side with the other side adopting a rectangular window. 
The TX window $U_{\nu}\left[ {n} \right]$ or the RX window $V_{\nu}\left[ {n} \right]$ can be obtained by equation (16) in \cite{Duhamel1953} according to the selected $\mathrm{SL}_w$.
In Fig. \ref{DifferentWindow}, for the same setting with Fig. \ref{FractionalIDI}, we employ a DC window at the transmitter side with $\mathrm{SL}_w [\mathrm{dB}]
 = -40$ dB and $k_{\mathrm{main}} \approx 3$.
We can observe that the resulting effective channel only has approximately $3$ entries with considerable gains and the channel spreading to all the other Doppler indices has been significantly suppressed due to the $40$ dB attenuation on the sidelobe  introduced by the designed DC window.
In other words, the adopted DC window can significantly suppress the channel spread and improve the effective channel sparsity, compared with the rectangular window.

\vspace{-4mm}
\section{Numerical Results}
In this section, we verify the accuracy of the derived analytical results and the effectiveness of the proposed designs via simulations.
We consider quadrature phase shift keying (QPSK) and binary phase shift keying (BPSK) modulations for the considered OTFS system.
For each OTFS frame, we set $N= [16,20]$ and $M = [8,30]$ indicating there are $N$ time slots and $M$ subcarriers in the TF domain.
The carrier frequency is $f_c = 3$ GHz and the subcarrier spacing is $\Delta f = 5$ kHz.
Without loss of generality, we set the maximum delay index as $l_{\mathrm{max}} = [2,4]$ and the maximum Doppler index as $k_{\mathrm{max}} = [2, 3]$.
The number of paths in the DD domain is $P = [2,5]$ and the additional guard space is $\hat k = [0,1]$.
For each channel realization, we randomly select the delay and Doppler indices such that we have $-k_{\mathrm{max}} \le {k_{{\nu _i}}} \le k_{\mathrm{max}}$ and $0 \le l_{\tau_i} \le l_{\mathrm{max}}$.
The channel coefficients $h_i$ are generated according to the distribution $h_{i}\sim \mathcal{CN}(0,q^{l_{\tau_i}})$, where $q^{l_{\tau_i}}$ follows a normalized exponential power delay profile\footnote{Here, we assume that the path loss and shadowing have been compensated following the literature studying OTFS \cite{RavitejaOTFS,YuanOTFS}.} $q^{l_{\tau_i}}=\frac{\exp(-0.1 l_{\tau_i})}{\sum_i \exp(-0.1 l_{\tau_i})}$.
We consider a normalized constellation $E\left\{ {{{\left| {x\left[ {k,l} \right]} \right|}^2}} \right\} = 1$, the system SNR is defined as $\mathrm{SNR} = \frac{1}{N_0}$, and the pilot power is $\left|x_p\right|^2 = [10,30]$ dBw\cite{RavitejaOTFSCE}.
The DC window is designed with $\mathrm{SL}_w [\mathrm{dB}] = -40$ dB such that $k_{\mathrm{main}} \approx 3$.
Two detectors including the MMSE detector in \eqref{MMSE} and the SPA-based detector \cite{Kschischang2001} are employed.
The key steps of SPA detection are briefly introduced in Appendix B for the sake of completeness.
All simulation results are averaged over more than $10^4$ OTFS frames.

\begin{figure}[t]
	\centering\vspace{-5mm}
	\includegraphics[width=3.8in]{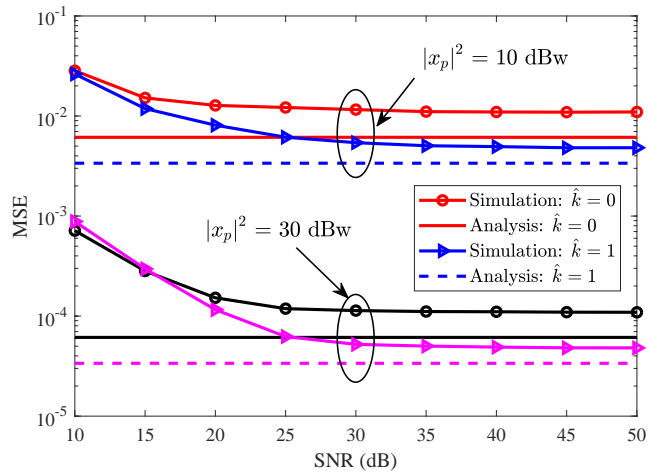}\vspace{-7mm}
	\caption{The MSE of effective channel estimation with employing rectangular window at the transceiver, where $N = 20$, $M = 30$, $k_{\mathrm{max}} = 3$, $l_{\mathrm{max}} = 4$, $P = 5$, $\hat k = [0,1]$, and $\left|x_p\right|^2 = [10,30]$ dBw. The corresponding relative speed between the transceiver is $270$ km/h.}\vspace{-10mm}
	\label{RectWindowInterspread}%
\end{figure}

\vspace{-4mm}
\subsection{Channel Estimation}
The effective channel estimation performance is evaluated in this section with rectangular and DC windows.
The results are shown in Fig. \ref{RectWindowInterspread}.
From the figure, we can observe that the effective channel estimation suffers from an error floor in all the considered cases.
This is due to the interference spread from data symbols to the guard space caused by the existence of the fractional Doppler.
Moreover, our derived error floor level in \eqref{ChannelESTIMATIONMSE} matches closely with the simulation results in the high SNR regime.
Note that a better effective channel estimation performance can be achieved with a higher pilot power.
Besides, as expected, the more additional guard space $\hat k$ inserted, the lower the MSE of channel estimation will be, at the expense of a higher amount of overhead.

\begin{figure}[t]
	\centering\vspace{-5mm}
	{\includegraphics[width=3.8in]{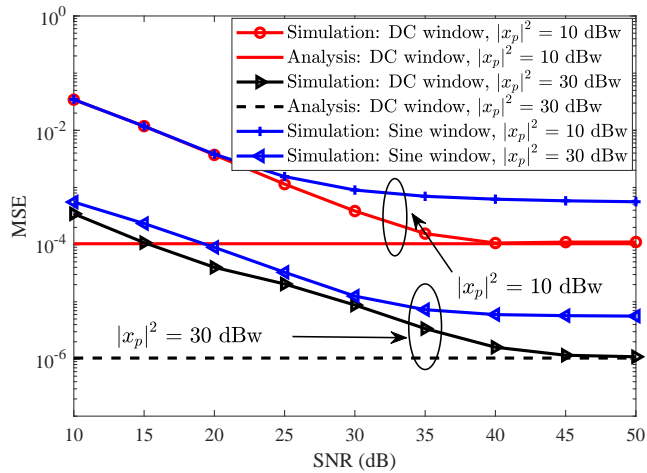}\vspace{-7mm}}
	\caption{The MSE of effective channel estimation with adopting DC window at the transmitter and rectangular window at the receiver, where $N = 20$, $M = 30$, $k_{\mathrm{max}} = 3$, $l_{\mathrm{max}} = 4$, $P = 5$, $\hat k = 1$, and $\left|x_p\right|^2 = [10,30]$ dBw. The corresponding relative speed between the transceiver is $270$ km/h.}\vspace{-5mm}
	\label{DCWindowChannelMSE}%
\end{figure}

Fig. \ref{DCWindowChannelMSE} illustrates the MSE of effective channel estimation when employing the designed DC window at the transmitter and rectangular window at the receiver.
We can observe that the derived error floor in \eqref{ChannelESTIMATIONMSE} is also consistent with the simulation results in the high SNR regime.
Comparing Fig. \ref{RectWindowInterspread} and Fig. \ref{DCWindowChannelMSE}, it can be seen that the employing the designed DC window at the transmitter is able to achieve a significantly lower MSE in the effective channel estimation than that of the rectangular window.
This demonstrates the effectiveness of the employed DC window in enhancing the effective channel sparsity and improving the effective channel estimation performance.
We also evaluate the MSE with employing the designed DC window at the receiver and a rectangular window at the transmitter.
The results are identical to those in Fig. 7.
This verifies that employing DC window at either the transmitter or the receiver can achieve the same error floor for the effective channel estimation in the high SNR regime, as predicted by our analysis.
On the other hand, since the channel estimation scheme in \eqref{DDCEFormula} does not exploit the statistics of the colored noise induced by employing DC window at the receiver, it achieves the same channel estimation performance as that of employing DC window at the transmitter in the medium-to-high SNR regime.

\vspace{-4mm}
\subsection{Data Detection}

\begin{table*}
	\vspace{0mm}
	\caption{Legend Label Explanation in Fig. \ref{BER_WoCSI} and Fig. \ref{BER_SPA}.}\vspace{-5mm}
	\centering
	\begin{tabular}{c|ccc}
		\hline
		Legend label     & Meaning \\\hline
		Rectangular  & Employing rectangular window at both the transmitter and the receiver \\\hline 
		DC TX        & Employing DC window at the transmitter and rectangular window at the receiver \\\hline
		DC RX        & Employing DC window at the receiver and rectangular window at the transmitter \\\hline 
		Optimized TX & Employing the proposed optimal TX window in \eqref{OptimalTxWindowMercury} \\\hline
		CSIR           & Perfect CSI at the receiver \\\hline
		CSIT \& CSIR   & Perfect CSI at both the transmitter and the receiver \\\hline
		Estimated CSIR & Estimated CSI at the receiver \\\hline
	\end{tabular}\label{LegendLabel}
\vspace{-10mm}
\end{table*}

\begin{figure}[t]
	\centering\vspace{-5mm}
	\includegraphics[width=3.8in]{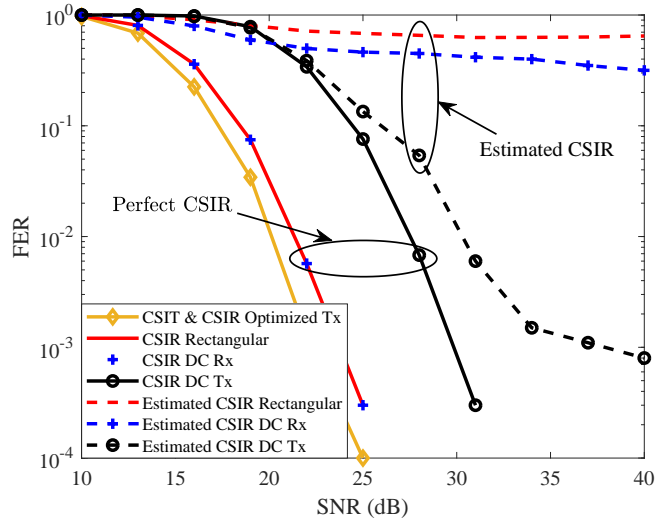}\vspace{-7mm}
	\caption{The FER performance with/without CSI employing different windows at the transmitter or receiver, where $N = 20$, $M = 30$, $k_{\mathrm{max}} = 3$, $l_{\mathrm{max}} = 4$, $P = 5$, $\hat k = 1$, and $\left|x_p\right|^2 = 30$ dBw. The corresponding relative speed between the transceiver $270$ km/h.}\vspace{-10mm}
	\label{BER_WoCSI}%
\end{figure}

The frame error rate (FER) performance of the MMSE detector is illustrated in Fig. \ref{BER_WoCSI}, where QPSK is adopted.
The legend labels are shown in Table \ref{LegendLabel}.
Firstly, we can observe that employing the proposed optimal TX window in \eqref{OptimalTxWindowMercury} (``CSIT \& CSIR Optimized Tx'') can achieve the best FER performance, compared to all the other cases.
This demonstrates the effectiveness of the proposed optimal window design in minimizing the detection MSE.
For the case with perfect CSI at the receiver, as predicted in \eqref{MMSERxTxWINDOW}, employing a DC window at the receiver (``CSIR DC RX'') does not affect the detection performance, resulting in an identical FER curve with that of the rectangular window (``CSIR Rectangular'').
Moreover, we can observe that employing a DC window at the transmitter (``CSIR DC TX'') yields a worse FER performance compared to that of the rectangular window (``CSIR Rectangular'').
This is because employing a DC window at the transmitter can enhance the channel sparsity while may sacrifice the detection MSE for the case with perfect CSIR.

For the case without CSI at the transmitter but CSI can be estimated at the receiver, we firstly employ the channel estimation scheme in \eqref{DDCEFormula} to estimate the effective channel in the DD domain and then perform MMSE detection in \eqref{MMSE} based on the estimated channels.
We can observe that due to the error floor of effective channel estimation as shown in Fig. \ref{RectWindowInterspread} and Fig. \ref{DCWindowChannelMSE}, the corresponding FER curves also exhibit error floors.
It can be seen that employing the DC window at the transmitter (``Estimated CSIR DC TX'') and the receiver (``Estimated CSIR DC RX'') can both achieve a lower error floor compared to that of rectangular window (``Estimated CSIR Rectangular'').
This is due to the enhanced channel sparsity and the improved channel estimation performance thanks to the proposed DC window.
Furthermore, employing the DC window at the transmitter can achieve a substantially lower FER error floor than that adopting it at the receiver.
In fact, as analyzed in \eqref{NoiseConvarianceMatrix}, introducing the DC window at the receiver results in colored noise while employing it at the transmitter still enjoys the white noise with identically enhanced channel sparsity.
The MMSE detection \eqref{MMSE} becomes more sensitive to the CSI imperfection for the case with colored noise, resulting in a poor detection performance with using the DC window at the receiver.


\begin{figure}[t]
	\centering\vspace{-2mm}
	\includegraphics[width=3.8in]{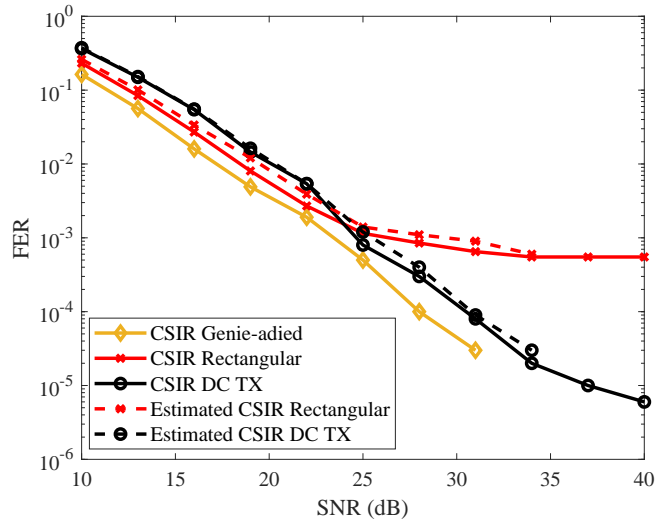}\vspace{-7mm}
	\caption{The FER performance based on SPA detection with/without CSI employing different windows at the transmitter or the receiver, where $N = 16$, $M = 8$, $k_{\mathrm{max}} = 2$, $l_{\mathrm{max}} = 2$, $P = 2$, $\hat k = 1$, and $\left|x_p\right|^2 = 30$ dBw. The corresponding relative speed between the transceiver is $112.5$ km/h.}\vspace{-10mm}
	\label{BER_SPA}%
\end{figure}

Now, Fig. \ref{BER_SPA} depicts the data detection performance with the classic sum-product algorithm (SPA)-based detector. 
Different from the MMSE detector, the complexity of SPA detector depends on the number of interfered symbols and the size of the constellation having a scale of $\mathcal{O}(M N {Q}^L)$, where $L$ denotes the number of interfered symbols. 
Thus, in the simulation case, BPSK is adopted with $Q = 2$.
Obviously, the reduction of $L$ would significantly reduce the receiver complexity.
Note that the number of interfered symbols $L$ is determined by the number of non-zero entries in the resulting effective channels in the DD domain.
Similar to \cite{RavitejaOTFS}, to have a fair comparison and to reduce the receiver complexity, we restrict the total number of interfered symbols to $L= 3 P-1$.
The residual interference due to the channel spread is treated as noise.
As adopting the DC window at the receiver suffers from the colored noise, we compare the performance with employing rectangular window at the transceiver and applying the proposed DC window at the transmitter.
Note that all these schemes have the same detection complexity since we assume the same number of interfered symbols $L$.
As a reference baseline, the FER performance employing rectangular window at the transceiver and a full-complexity SPA detector with CSIR (``CSIR Genie-aided'') is illustrated, which serves as a lower bound of FER for all the other cases.
It can be seen that the detection performance for the case of estimated CSIR corresponding to both the DC window and the rectangular window can attain their perfect CSIR counterparts, which verifies the effectiveness of the channel estimation scheme in \eqref{DDCEFormula}.
We can see that employing the DC window at the transmitter side will slightly degrade the performance in the low SNR regime compared with that of the rectangular window.
This is because in the low SNR regime, the data detection performance is mainly limited by the noise instead of the channel estimation error.
Therefore, improving the effective channel sparsity via the designed DC window does not yield a significant gain in data detection in the low SNR regime.
However, an error floor can be observed for the case adopting rectangular window, while the DC window does not lead to such an error floor until $\mathrm{SNR} = 40$ dB.
In fact, as shown in Section IV, the proposed DC window can improve the sparsity of the effective DD domain channel and thus restricting $L$ causes a smaller residual interference and does not lose too much information.
In contrast, due to the channel spread for the case of rectangular window, a low-complexity detection with a finite $L$ would suffer from a higher residual interference and cause a severe information loss, despite the availability of the CSIR.

\vspace{-4mm}
\section{Conclusions}
In this paper, we investigated the impacts of transmitter and receiver windows and proposed window designs for OTFS modulation.
We analyzed and revealed the insights of the impacts of windowing on the effective channel, the effective channel estimation performance, the total average transmit power, and the noise covariance matrix.
In particular, we showed that the existence of fractional Doppler leads to the potential effective channel spread, which causes an error floor in the effective channel estimation. 
We found that adopting a window at the transmitter or the receiver can obtain an identical performance in the effective channel estimation.
Besides, the TX window can be interpreted as the power allocation in the TF domain, while employing a RX window causes colored noise.
When CSI is available at both the transmitter and receiver, we analyzed the data detection MSE adopting an MMSE detector and proposed an optimal TX window to minimize the detection MSE.
When CSI is not available at the transmitter but can be estimated at the receiver, we proposed to apply a DC window to enhance the channel sparsity, which improves the performance in both channel estimation and data detection.
We verified the accuracy of the obtained analytical results and insights and demonstrated the substantial performance gain of the proposed window designs.

\vspace{-4mm}
\section*{Appendix}

\vspace{-4mm}
\subsection{Sum-product Algorithm (SPA)-based Detector\cite{Kschischang2001}}\label{SPA_detection}
The SPA detection can be carried out based on the optimal maximum \emph{a posteriori} (MAP) detection, i.e.,
\vspace{-4mm}
\begin{equation}
\hat{x}\left[k,l\right] = \arg\max_{{x}\left[k,l\right]} p({x}\left[k,l\right]|\mathbf{y}_{\rm DD}),\vspace{-2mm}
\end{equation}
where $p({x}\left[k,l\right]|\mathbf{y}_{\rm DD})$ denotes the \emph{a posteriori} probability.
Due to the interference channel, each data symbol ${x}\left[k,l\right]$ in a received sample ${y}\left[\bar{k},\bar{l}\right]$ will be interfered by other data symbols.
For the case of integer Doppler, it is easy to find that each data symbol is interfered by $P-1$ symbols for a given received sample.
For the case of fractional Doppler, channel spread drastically increases the number of interfered symbols, depending on the number of non-zero entries in the resulting effective channels in the DD domain.
Similar to \cite{RavitejaOTFS}, to reduce the computational complexity, we only select $L$-largest entries of the effective channels to account the signals contributed most and all the other signals can been seen as noise.
For the ease of exposition, we assume that each data symbol is interfered by $L$ symbols given a received sample. In turn, a data symbol ${x}\left[k,l\right]$ will appear in $L$ received samples in an OTFS frame.
Let us denote the set of all received samples containing ${x}\left[k,l\right]$ by $\mathbb{Y}_{k,l}$, whose $i$th element is $\mathbb{Y}^{i}_{k,l}$. We further denote $\mathbb{X}_{k,l}^{i}$ as the set of all data symbols corresponding to $\mathbb{Y}^{i}_{k,l}$ except ${x}\left[k,l\right]$. Then the SPA detection is given by
\vspace{-2mm}
\begin{align}\label{sumstep}
&{\rm Sum~step:}~ p(\mathbb{Y}^{i}_{k,l}\to {x}\left[k,l\right] ) = \sum_{\mathbb{X}_{k,l}^{i}} p(\mathbb{Y}^{i}_{k,l}|{x}\left[k,l\right],\mathbb{X}_{k,l}^{i}) \prod_{x\left[k',l'\right]\in \mathbb{X}_{k,l}^{i}}p (x\left[k',l'\right]\to \mathbb{Y}^{i}_{k,l}),\\[-1mm]
&{\rm Product~step:}~p({x}\left[k,l\right]\to\mathbb{Y}^{i}_{k,l})  = p({x}\left[k,l\right])\prod_{i'\neq i }p(\mathbb{Y}^{i'}_{k,l}\to {x}\left[k,l\right] ).
\end{align}
\vspace{-8mm}\par\noindent
The probabilities are updated iteratively following the sum and product steps. Finally, the \emph{a posteriori} probability $p({x}\left[k,l\right]|\mathbf{y}_{\rm DD})$ is given by
\vspace{-2mm}
\begin{equation}
p({x}\left[k,l\right]|\mathbf{y}_{\rm DD}) = p({x}\left[k,l\right]|\mathbb{Y}_{k,l}) = p({x}\left[k,l\right])\prod_{i}p(\mathbb{Y}^{i}_{k,l}\to {x}\left[k,l\right] ),\vspace{-2mm}
\end{equation}
which is used for detecting the data symbol ${x}\left[k,l\right]$. Note that the complexity of SPA detection is dominated by the sum step \eqref{sumstep}. Given the size of $\mathbb{X}_{k,l}^{i}$ being $L$ and the constellation size $Q$, the summation in \eqref{sumstep} has to be performed $Q^L$ times, leading to a complexity scale of $\mathcal{O}(Q^L)$.

\bibliographystyle{IEEEtran}
\bibliography{OTFS}

\end{document}